\documentclass[rmp,amsfonts,amssymb,nofootinbib,twocolumn,amsmath,superscriptaddress]{revtex4-2}

\usepackage{graphicx}
\usepackage{bm}
\usepackage{hyperref}

\setcitestyle{numbers,sort&compress}           
\usepackage{xcolor,amsmath}

\begin{document}

\title{Many-body quantum geometric effects and entanglement at the 3D metal-insulator quantum phase transition}

\author{Jason Y. Yan}
\affiliation{William H. Miller III Department of Physics and Astronomy, The Johns Hopkins University, Baltimore, Maryland 21218, USA}

\author{Xiaoyu Guo}
\affiliation{William H. Miller III Department of Physics and Astronomy, The Johns Hopkins University, Baltimore, Maryland 21218, USA}

\author{R. Bhandia}
\affiliation{William H. Miller III Department of Physics and Astronomy, The Johns Hopkins University, Baltimore, Maryland 21218, USA}

\author{T.F. Rosenbaum}
\affiliation{Division of Physics, Mathematics, and Astronomy, California Institute of Technology, Pasadena, CA 91125, USA}

\author{Natalia Drichko}
\affiliation{William H. Miller III Department of Physics and Astronomy, The Johns Hopkins University, Baltimore, Maryland 21218, USA}

\author{N. P. Armitage}\email{npa@jhu.edu}
\affiliation{William H. Miller III Department of Physics and Astronomy, The Johns Hopkins University, Baltimore, Maryland 21218, USA}

\date{\today}

\maketitle

\textbf{ Quantum geometry has emerged as a unifying concept across condensed matter physics~\cite{provost1980riemannian,torma2023essay,yu2025quantum,verma2026quantum}, underlying phenomena from nonlinear topological response~\cite{morimoto2016topological} to flat-band superconductivity~\cite{peotta2015superfluidity}. While usually formulated within band theory, quantum geometry remains meaningful in disordered interacting systems~\cite{resta1999electron}. Here we show that the first negative moment of the optical conductivity~\cite{Kudinov91a,souza2000polarization,verma2025instantaneous,souza2025optical,onishi2025quantum} -- proportional to the zero temperature quantum Fisher information as a bound on the multipartite entanglement~\cite{hauke2016measuring,scheie2021witnessing} -- provides an experimental probe of quantum geometry across the three-dimensional metal-insulator quantum phase transition in phosphorus-doped silicon.  We extract a quantum geometric length $\ell$ that characterizes the local wavefunctions. Far from the transition, this length is almost coincident with the Bohr radius of the hydrogenic phosphorus donors, reflecting their atomic-scale quantum geometry. Approaching the transition, $\ell$ is enhanced, but does not diverge continuously like a correlation length; it jumps discontinuously to infinity at the critical point.  This reflects the UV domination of the sum rule in three dimensions that renders it insensitive to the critical fluctuations driving the diverging dielectric constant and correlation length.  Its enhancement demonstrates a ``puffing" of the donor polarizability volume of quantum geometric origin, which yields a quantum geometric corrected Clausius-Mossotti description in closer agreement with the diverging dielectric response and provides a quantum mechanical foundation for the century-old Herzfeld metallization criterion~\cite{herzfeld1927atomic}.}

\bigskip

The first negative moment sum rule for the optical conductivity,  -- known since the 1990s~\cite{Kudinov91a,souza2000polarization} (and in specific contexts even earlier~\cite{brauwers1975sum,kivelson1982wannier}) is

\begin{align}
\label{SWM}
S^{ab}_{-1} =  \frac{2 e^2 n_e}{\hbar} \langle \ell_a \ell_b \rangle =  \frac{2}{\pi} \int_{0}^{\infty} \frac{\mathrm{Re}\,\sigma_{ab}(\omega)}{\omega} \, d\omega \nonumber \\ =   \frac{2}{\pi} \int_{0}^{\infty} \chi"(\omega)  \, d\omega = \frac{F_Q(T \rightarrow 0)}{2}  ,
\end{align}
where $n_e$ is the density of electrons, $ \langle \ell_a \ell_b \rangle $ is the localization tensor (the second cumulant moment of the electron distribution), and $\chi"$ is the dissipative part of susceptibility.  This relation follows from the correspondence between susceptibility and conductivity ($\sigma = i \omega \epsilon_0 \chi$).  $F_Q$ is the quantum Fisher information, which bounds multipartite entanglement~\cite{hauke2016measuring}.  $S^{ab}_{-1}$ was shown by Souza, Wilkens, and Martin (SWM) to be a measure of the ground-state quantum metric and directly proportional to the scale of zero point polarization fluctuations~\cite{souza2000polarization}.  The relation has been under exploited in condensed matter physics.  The diagonal entries of the localization tensor $\langle \ell_a \ell_b \rangle $ carry the interpretation of a localization length squared and are expected to measure how spatially spread the many-electron ground state is.  In non-interacting band models, $S_{-1}$ provides a gauge-invariant measure of the minimum spatial spread of the Wannier single-particle wave functions~\cite{marzari1997maximally} and the trace of the quantum metric tensor, thereby offering a purely geometric characterization of electronic states.  For an ensemble of isolated hydrogen atoms, the SWM sum rule applied to $\sigma_{xx}$ is predicted to return $\ell^2 = a^2$~\cite{souza2025optical,traini1996electric} where $a$ is the Bohr radius.  The sum rule remains well-defined in interacting and disordered systems, where it takes on the role of a many-body localization length~\cite{resta1999electron,mao2025low,ozawa2019probing,faugno2026characterizing}.  In fact the band quantum geometry relevant to translationally invariant systems is a special case of a more general many-body, quantum geometry that can be naturally formulated in real space~\cite{resta1999electron}.  In a translationally invariant system, position matrix elements become the interband Berry connection in momentum space that can be used to build the band quantum geometry~\cite{souza2000polarization}.

The $S_{-1}$ sum rule has been used in a different context to quantify the quantum Fisher information from the magnetic and x-ray scattering~\cite{scheie2021witnessing,balut2025quantum,balut2026fundamental} and multi-partite entanglement.  Similar to the notion of a Wannier function there, it gives the minimal size of a local quantum mechanically entangled object that larger extended states can be built from.  A measure of the quantum geometric tensor has recently been performed in the kagome metal CoSn by angle-resolved photoemission spectroscopy~\cite{kang2025measurements}, but it would be illuminating to see how it affects macroscopic observables particularly as one tunes through a quantum phase transition.

\begin{figure}[tbp]
\includegraphics[width=0.48\textwidth]{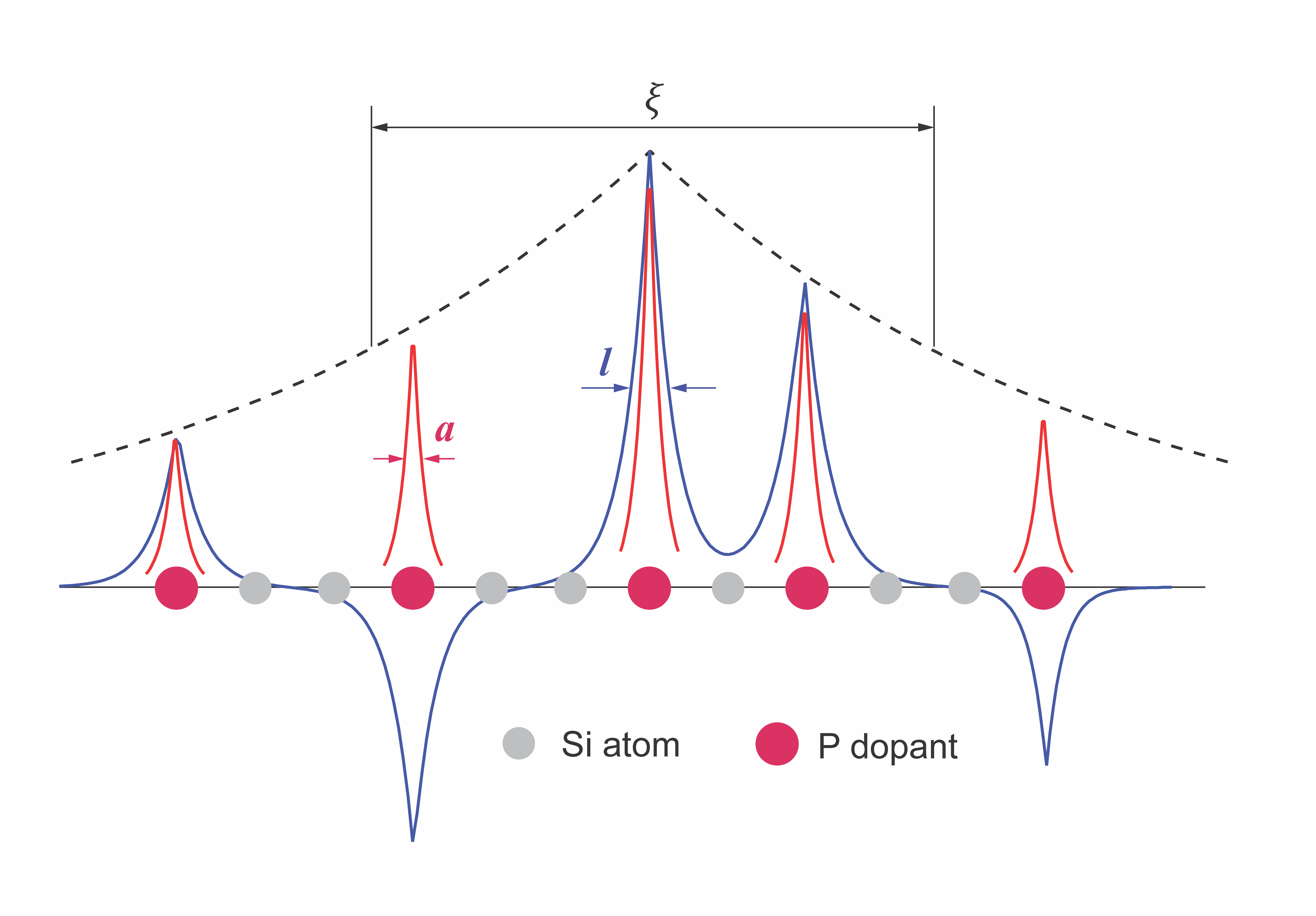}
\caption{{\bf Schematic of wavefunctions of a non-interacting Anderson insulator.} $\xi$ is the overall localization length of the wavefunction that is made by the hybridization of localized states.   $\ell$ is the quantum geometric length, which is the size of the local states.  It may be distinguished from $a$ which is the size of an isolated orbital (the Bohr radius for a hydrogenic wavefunction).  }
\label{Wavefunctions} 
\end{figure}

Phosphorus-doped silicon (Si:P) is a model system with a zero temperature metal-insulator quantum phase transition (MIQPT) as a function of the phosphorus concentration~\cite{capizzi1980observation,rosenbaum1983metal,Helgren02a,Helgren04a,hering2007signature}.  It is widely studied as a model system for electron glass physics on the insulating side of its transition.   At low concentrations, deep in the insulator one can model the n-type dopants as giving isolated randomly located hydrogenic orbitals of modified atomic parameters~\cite{Gayman93a,Thomas81a}.  State-of-the-art {\it ab initio} calculations give an effective Bohr radius of about $a = 1.8$ nm~\cite{smith2017ab}, which is near the textbook Bohr model prediction.  At increased concentrations, these orbitals start to overlap, the low frequency dielectric constant becomes greatly enhanced, and when the doping concentration reaches $x_c \cong 3.5\times10^{18}$/cm$^3$ there is a MIQPT to a conducting state~\cite{capizzi1980observation,rosenbaum1983metal,Helgren02a,Helgren04a,hering2007signature}.  The dielectric constant diverges at the MIQPT, reflecting a diverging correlation length $\xi$~\cite{mcmillan1981scaling} that in the limit of a non-interacting Anderson insulator coincides with the scale of the exponential envelope of a localized wavefunction at the Fermi energy (Fig.~\ref{Wavefunctions}). Therefore, there are three emerging length scales: $\xi$ measures the size of the overall wavefunction cluster size; $\ell$ measures the size of local states; $a$ measures the Bohr radius of individual P dopants.  The central result of this work is that $\xi$ and $\ell$ are distinct length scales near the MIQPT, exhibiting different doping dependencies and characteristic scales relative to $a$.

 \begin{figure}[tb]
    \includegraphics[width=8.9cm]{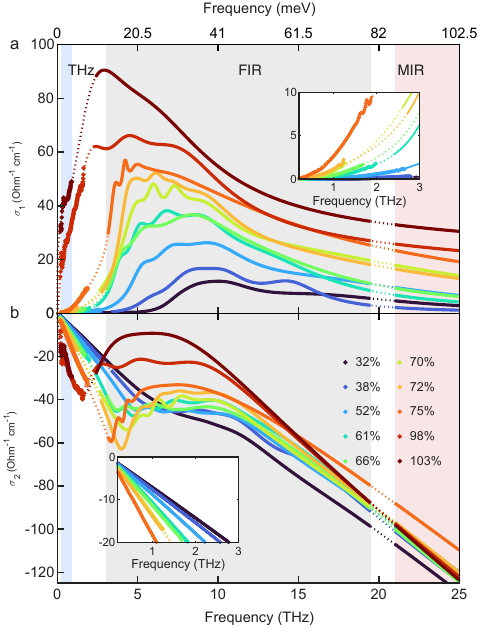}
    \caption{{\bf Optical conductivity from THz and FTIR spectroscopy}. (a) Real part of the optical conductivity $\sigma_1(\omega)$ as a function of frequency at the dopings indicated.  (b) Imaginary part of conductivity $\sigma_2(\omega)$ as a function of frequency at the same dopings. Data shown as thin dashed line is interpolated as discussed in the text.  Insets show the low frequency regimes. Shaded regions indicate the different experimental frequency ranges. (See Methods for further details).}
    \label{fig:conductivity}
\end{figure}

In Fig.~\ref{fig:conductivity}, we present the real and imaginary parts of the optical conductivity obtained from combining time-domain THz spectroscopy (TDTS) transmission and Fourier transform infrared (FTIR) reflectivity measurements.  TDTS returns the complex conductivity directly by numerical inversion of the measured complex transmission function.  The complex conductivity over the full range was obtained by a global fitting in a Kramers-Kronig consistent manner of both the real FTIR reflectivity data and complex TDTS transmission data to a dense ensemble of Lorentz oscillators as described in the Methods and Ref.~\cite{kuzmenko2005kramers}.  At much lower dopings than shown in Fig.~\ref{fig:conductivity} (See SM Fig.~S\ref{Thomas}), the conductivity shows a number of very sharp peaks reflecting transitions between well-defined hydrogenic-like orbitals and a transition to the conduction band continuum at approximately 43 meV~\cite{Thomas81a,Gayman93a}.  As the dopant density increases, these states broaden and the continuum peak moves down in energy and overlaps with the bound-state resonances.  Peak broadening is typically asymmetric towards lower energies, and has distinct features corresponding to the median sized clusters~\cite{Thomas81a}.  As the MIQPT is approached a low frequency power law regime is found in $\sigma_1$(see inset to Fig.~\ref{fig:conductivity}) as previously discussed~\cite{Shklovskii81a,Lee01a,Helgren02a,Helgren04a,hering2007signature}.  The MIQPT occurs at the concentrations above which $\sigma_1(\omega \rightarrow 0, T\rightarrow 0)$ becomes finite.  At 103\% $x_c$, the system exhibits dirty metallic behavior, characterized by a finite-frequency peak and nonzero DC conductivity.

\begin{figure}
    \centering
\includegraphics[width=0.95\linewidth]{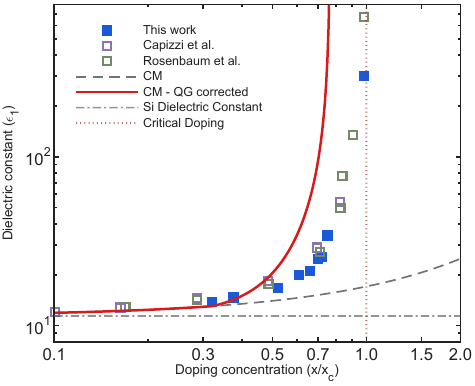}
    \caption{{\bf Dielectric constant of Si:P.} The measured low frequency dielectric from the TDTS experiments as a function of dopant concentration.   We overlay historical data from Refs.~\cite{rosenbaum1983metal,capizzi1980observation} which except for the 98$\%$ shows consistency with our data.   The gray dashed line is fit to the Clausius-Mossotti (CM) expression using the known experimental measured single dopant polarizability $\alpha_D = 110\;$nm$^3$.   The red line is a {\it quantum geometry corrected} CM expression where we have scaled $\alpha_D$ by a ratio of $\big[\ell(x)/\ell(0\big]^3$ using the fit in Fig.~\ref{lengths}.  }
    \label{dielectric}
\end{figure}

From Fig.~\ref{fig:conductivity}b, the low-frequency region of the imaginary conductivity is linear in frequency, allowing the almost DC dielectric constant to be determined from the relation $\epsilon_1 = 1 +  \frac{\sigma_2}{\varepsilon_0 \omega}$.  In Fig.~\ref{dielectric}, we plot this low frequency dielectric constant as a function of doping.  At densities far from the transition the dielectric constant is described by the simple relation $\epsilon = \epsilon_{si}  +  4 \pi x \alpha_D $, where $\alpha_D$ is the polarizability of a single phosphorus dopant, $\epsilon_{si}$ is the host material dielectric constant, and $x$ is the dopant concentration.  For a hydrogenic orbital in silicon $\alpha_D = \frac{9}{2} \epsilon_{si} a^3 = 299\;$nm$^3$ using $a=1.8\;$nm and $\epsilon_{si} = 11.4$.  In fact, a fit to the low doping region reveals a smaller $\alpha_D$ of approximately $110\;$nm$^3$, which is a large enough difference that it may indicate deviations from an ideal hydrogenic wavefunction not captured by just a smaller $a$.  As the critical concentration is approached the dielectric constant diverges in a fashion consistent with earlier work~\cite{capizzi1980observation,rosenbaum1983metal,tan1981piezocapacitance}.  The discrepancy between the THz data and the data point at 98$\%$ from Ref.~\cite{rosenbaum1983metal} is likely because the latter point was taken at much lower frequencies using a 400 MHz resonant transmission cavity, which becomes necessary as spectral weight shifts towards zero on the approach to the MIQPT.  As can be seen from the gray dashed line in Fig.~\ref{dielectric}, a correction to the linear dependence from the Clausius-Mossotti relation $\epsilon = \epsilon_{si}  + \frac{4 \pi x \alpha_D}{ 1 - 4 \pi x \alpha_D/3\epsilon_{si} }$ accounts for the nonlinear upward trend of $\epsilon$, but as shown earlier does {\it not} describe the divergence of dielectric constant correctly~\cite{castner1980dielectric,tan1981piezocapacitance}.  It predicts from the Herzfeld condition~\cite{herzfeld1927atomic} a critical concentration of 2.4$\times 10^{19}/$cm$^3$, which is almost an order of magnitude larger than observed.  

\begin{figure}
    \centering
    \includegraphics[width=0.95\linewidth]{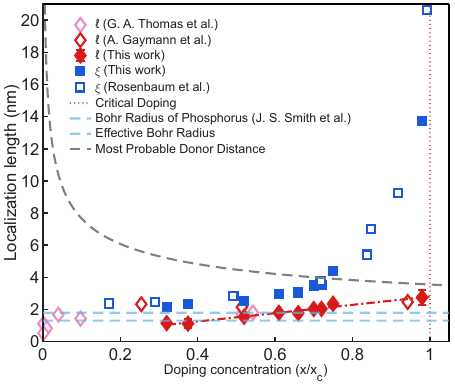}
    \caption{{\bf Quantum geometric length $\ell$ and correlation length $\xi$ as a function of doping.}  
    $\xi$ (blue square) diverges approaching the MIQPT while $\ell$ (red diamond) does not. The upper horizontal line is the calculated Bohr radius~\cite{smith2017ab}.  The lower horizontal line is a length scale $a_{eff}$ determined by the assumption of a hydrogenic form and the experimental value of the isolated dopant polarizability $\alpha_D = \frac{9}{2} \epsilon_{si} a^3_{eff}$.  Also plotted is the most probable nearest neighbor distance between dopants based on a Poisson distribution e.g. $\langle r_{nn} \rangle = 0.54 x^{-1/3} $.  Historical data from Refs.~\cite{Gayman93a} and \cite{Thomas81a} are overlayed as open symbols. The red dashed line is a linear fit giving general trend of $\ell$, and error bars for $\ell$ are determined from the conductivity.   Data on the far left from Ref.~\cite{Thomas81a} is not zero but as low as $x=4.4 \times10^{15}$/cm$^3$.   }
    \label{lengths}
\end{figure}

We now move on to discuss the quantum geometric length $\ell$ on the approach to the MIQPT. We use Eq.~\ref{SWM} and the data in Fig.~\ref{fig:conductivity}a to evaluate the first negative moment sum rule of the conductivity.  The integral was performed by using the fitted spectral weights thereby incorporating to relatively high fidelity the contribution of the high frequency tail (see SM Sec. VI) of Fig.~\ref{fig:conductivity}a as well as the relatively small amount of spectral weight below the measurement range.  We plot the quantum geometric length $\ell$ in Fig.~\ref{lengths} vs. the dopant concentration\footnote{Formally the SWM integral should be performed over all frequencies. In the present case where $\ell \gg$ 0.17 nm (the Wannier spread size in Si) one incurs corrections only of order 0.05$\%$ to the integral by not integrating over the energy of valence to conduction band excitations~\cite{souza2025optical}}.  One can see that for very dilute samples the integral returns a length scale $\ell$ of the order of the accepted Bohr radius of P in Si. Interestingly, at low dopings $\ell$ is closer to the length set by inverting the hydrogenic form for the dopant susceptibility using the measured $\alpha_D$ e.g. $a_{eff} = \sqrt[3]{\frac{2 \alpha_D}{9 \epsilon_{si} }} =  1.3\;$nm, again perhaps pointing to deviations from a pure hydrogenic wavefunction.  Over the full measured range, $\ell$ increases by about a factor of 2.4, but stays finite right up to the critical concentration.  This behavior can be contrasted to the correlation length $\xi$ that characterizes the MIQPT.  One expects~\cite{mcmillan1981scaling} a scaling relation $\epsilon =  \epsilon_{Si} + \epsilon_a  \big(\frac{\xi}{a}\big)^{z - (d-2)}$ (where $ \epsilon_a = 4 \pi x \alpha_D $).   Choosing the uncorrelated value of $z=d$, which is consistent with earlier work~\cite{rosenbaum1983metal} and using the data of Fig.~\ref{dielectric} we plot $\xi$ in Fig.~\ref{lengths}.   One can see $\xi$ strongly diverges as the MIQPT is approached in dramatic contrast to $\ell$.  Note that since we have chosen what is presumably the largest possible value for $z$ of $z=d$ our estimate for $\xi$ is in fact a lower bound on the actual correlation length.

As discussed in the SM Sec. IV, using constraints from the SWM sum rule, the $f$-sum rule, and the Kramers-Kronig relations for the dielectric constant, one can derive a bound  $\frac{3}{2 \sqrt{2}}  \frac{\xi}{a} \geq \left (\frac{\ell}{a} \right )^2 $ obeyed by our data.   As discussed in SM Sec. IV and Ref.~\cite{souza2025optical}, this is almost an equality for hydrogenic orbitals.   The large difference between $\xi$ and $\ell$ imply that estimates of the quantum geometry from the dielectric constant~\cite{souza2025optical,komissarov2024quantum,onishi2025quantum,verma2025instantaneous} are likely to be inaccurate when the spectra departs significantly from a single mode shape as is of course the case near the MIQPT.  The differences in $\ell$ and $\xi$ are supported by our scaling analysis (see SM Sec. VI) that shows that any doping dependence of $\ell$ is unrelated to the divergence in the MIQPT correlation length $\xi$.  This is because in comparison to the the dielectric constant the SWM integral in 3D is dominated by UV-scale optical weight, so it is insensitive to the critical softening that drives the transition.  In contrast to what has been inferred from the 1D Mott transition and despite the fact that it is finite in the insulator and infinite in the metal, $\ell$ {\it does not} play the role of a continuously diverging critical length for the 3D MIQPT.  It jumps discontinuously to infinity at $x_c$.  Although no understanding of a full many-body disordered interacting system exists, the case of the non-interacting Anderson insulator is illuminating as shown in Fig.~\ref{Wavefunctions}.  In such a case, $\xi$ is the length scale of the envelope of the single particle wavefunction at E$_F$ and $\ell$ is the scale of the local polarizable object, which is in contrast to $a$ which is the size of an isolated  orbital.   

What is the significance of the non-divergent enhancement of $\ell$ with increased doping?   We propose that its increase can be interpreted as an enhancement of the polarizability volume e.g. a ``puffing" of the orbital size on the approach to the MIQPT that presumably arises from the renormalizing effects of intersite dipole fluctuations.  To verify this we scale $\alpha_D$ by the ratio of $\big [ \frac{\ell(x)}{\ell(0) }\big ]^3$ and plot a ``quantum geometry corrected" Clausius-Mossotti expression in Fig.~\ref{dielectric}.  As seen, this changes the predicted $x_c$ from being about an order of magnitude too large to being about 20$\%$ of the observed value.  Of course a mean-field model like Clausius-Mossotti cannot be the whole story, but it is remarkable that with quantum geometric corrections it gets the approximate scale of $x_c$ roughly correct.  In a related fashion, it may be interesting to formulate an entanglement-based condition for the occurrence of the MIQPT based on the relationship between the SWM sum rule and the QFI discussed above.  Note that the QFI is not itself a measure of entanglement -- its dominant part is the single-particle orbital response, a separable contribution present already for independent donors -- but its inter-site, non-separable component is the cooperative enhancement and reflects entanglement between donors.   We give an example in the SM Sec. VII.

Our results demonstrate that the first negative moment sum rule of the conductivity offers a robust and experimentally accessible route to characterizing ground state properties  of interacting disordered systems.  We find the quantum geometric length $\ell$ is enhanced -- presumably from intersite dipole fluctuations arising from virtual transitions between neighboring donors, but does not diverge on the approach to the MIQPT.  $\ell$ {\it is} simultaneously the length scale of polarization fluctuations, an effective Clausius-Mossotti radius, and a measure of quantum information content of the ground state with respect to electromagnetic perturbations.  It is {\it not} the critical length of the MIQPT, highlighting the open question of what does play this role in an interacting disordered system as such.  Our work unifies insights from quantum geometry, many-body localization, and quantum information, extending the applicability of geometric approaches from translationally invariant models to the complex, correlated materials often found in the laboratory.

\bigskip

{\bf Methods.} Experiments were performed on nominally uncompensated phosphorus-doped Si (Si:P) samples.  Samples of 32-75$\%$ of the critical concentration of the 3D MIQPT were cut from a Czochralski grown boule of 5 cm in diameter with a P-dopant gradient along the axis. This boule was sliced and  polished down to 100 $\mu$m wafers. Samples from this boule were previously used for studies of the THz-range conductivity in the phononless regime, optical pump-THz probe measurements, and 2D THz coherent spectroscopy \cite{Helgren04a,Helgren02a,thorsmolle2010ultrafast,mahmood2021observation}. Samples of 98$\%$ and 103$\%$ doping were previously measured in Refs.~\cite{rosenbaum1983metal,Thomas81a}.

Time-domain THz spectroscopy (TDTS) from 0.2 - 2 THz (0.83 - 8.3 meV) was performed in a standard transmission configuration to obtain the complex transmission function, which was then be inverted by numerical Newton-Raphson to get the complex conductivity.  For frequencies above those possible with TDTS we performed Fourier transform infrared (FTIR) reflectivity measurements on a Bruker 80v.  For frequencies in the 3 - 19.5 THz (FIR 12.4 - 80.6 meV) range we used a Hg lamp and Helium cooled bolometer in reflection.  In the 21 - 210 THz (MIR 86.8 - 868.5 meV) range we performed reflectivity measurements using a glow-bar and an MCT detector. Reflectivity was referenced to evaporated gold. Both TDTS and FTIR experiments were measured in the zero temperature limit ($\hbar \omega \gg kT$) where there is minimal temperature dependence. We combined the complex THz conductivity data with the reflectivity in a Kramers-Kronig consistent variational fitting regime to get the complex conductivity data over a large frequency range~\cite{kuzmenko2005kramers}.  Through the constraints afforded to the fitting due to Kramers-Kronig compatibility, such analysis has been shown to provide reliable interpolation even for small frequency ranges that data was not taken in as is the case here.

\bigskip

\textbf{Data availability.} All data that support the plots within this paper are available from the corresponding author upon reasonable request

 \bibliography{EGlassBib}

\begin{thebibliography}{49}%
\makeatletter
\providecommand \@ifxundefined [1]{%
 \@ifx{#1\undefined}
}%
\providecommand \@ifnum [1]{%
 \ifnum #1\expandafter \@firstoftwo
 \else \expandafter \@secondoftwo
 \fi
}%
\providecommand \@ifx [1]{%
 \ifx #1\expandafter \@firstoftwo
 \else \expandafter \@secondoftwo
 \fi
}%
\providecommand \natexlab [1]{#1}%
\providecommand \enquote  [1]{``#1''}%
\providecommand \bibnamefont  [1]{#1}%
\providecommand \bibfnamefont [1]{#1}%
\providecommand \citenamefont [1]{#1}%
\providecommand \href@noop [0]{\@secondoftwo}%
\providecommand \href [0]{\begingroup \@sanitize@url \@href}%
\providecommand \@href[1]{\@@startlink{#1}\@@href}%
\providecommand \@@href[1]{\endgroup#1\@@endlink}%
\providecommand \@sanitize@url [0]{\catcode `\\12\catcode `\$12\catcode
  `\&12\catcode `\#12\catcode `\^12\catcode `\_12\catcode `\%12\relax}%
\providecommand \@@startlink[1]{}%
\providecommand \@@endlink[0]{}%
\providecommand \url  [0]{\begingroup\@sanitize@url \@url }%
\providecommand \@url [1]{\endgroup\@href {#1}{\urlprefix }}%
\providecommand \urlprefix  [0]{URL }%
\providecommand \Eprint [0]{\href }%
\providecommand \doibase [0]{https://doi.org/}%
\providecommand \selectlanguage [0]{\@gobble}%
\providecommand \bibinfo  [0]{\@secondoftwo}%
\providecommand \bibfield  [0]{\@secondoftwo}%
\providecommand \translation [1]{[#1]}%
\providecommand \BibitemOpen [0]{}%
\providecommand \bibitemStop [0]{}%
\providecommand \bibitemNoStop [0]{.\EOS\space}%
\providecommand \EOS [0]{\spacefactor3000\relax}%
\providecommand \BibitemShut  [1]{\csname bibitem#1\endcsname}%
\let\auto@bib@innerbib\@empty
\bibitem [{\citenamefont {Provost}\ and\ \citenamefont
  {Vallee}(1980)}]{provost1980riemannian}%
  \BibitemOpen
  \bibfield  {author} {\bibinfo {author} {\bibfnamefont {J.}~\bibnamefont
  {Provost}}\ and\ \bibinfo {author} {\bibfnamefont {G.}~\bibnamefont
  {Vallee}},\ }\href@noop {} {\bibfield  {journal} {\bibinfo  {journal}
  {Communications in Mathematical Physics}\ }\textbf {\bibinfo {volume} {76}},\
  \bibinfo {pages} {289} (\bibinfo {year} {1980})}\BibitemShut {NoStop}%
\bibitem [{\citenamefont {T{\"o}rm{\"a}}(2023)}]{torma2023essay}%
  \BibitemOpen
  \bibfield  {author} {\bibinfo {author} {\bibfnamefont {P.}~\bibnamefont
  {T{\"o}rm{\"a}}},\ }\href@noop {} {\bibfield  {journal} {\bibinfo  {journal}
  {Physical Review Letters}\ }\textbf {\bibinfo {volume} {131}},\ \bibinfo
  {pages} {240001} (\bibinfo {year} {2023})}\BibitemShut {NoStop}%
\bibitem [{\citenamefont {Yu}\ \emph {et~al.}(2025)\citenamefont {Yu},
  \citenamefont {Bernevig}, \citenamefont {Queiroz}, \citenamefont {Rossi},
  \citenamefont {T{\"o}rm{\"a}},\ and\ \citenamefont {Yang}}]{yu2025quantum}%
  \BibitemOpen
  \bibfield  {author} {\bibinfo {author} {\bibfnamefont {J.}~\bibnamefont
  {Yu}}, \bibinfo {author} {\bibfnamefont {B.~A.}\ \bibnamefont {Bernevig}},
  \bibinfo {author} {\bibfnamefont {R.}~\bibnamefont {Queiroz}}, \bibinfo
  {author} {\bibfnamefont {E.}~\bibnamefont {Rossi}}, \bibinfo {author}
  {\bibfnamefont {P.}~\bibnamefont {T{\"o}rm{\"a}}},\ and\ \bibinfo {author}
  {\bibfnamefont {B.-J.}\ \bibnamefont {Yang}},\ }\href@noop {} {\bibfield
  {journal} {\bibinfo  {journal} {npj Quantum Materials}\ }\textbf {\bibinfo
  {volume} {10}},\ \bibinfo {pages} {101} (\bibinfo {year} {2025})}\BibitemShut
  {NoStop}%
\bibitem [{\citenamefont {Verma}\ \emph {et~al.}(2026)\citenamefont {Verma},
  \citenamefont {Moll}, \citenamefont {Holder},\ and\ \citenamefont
  {Queiroz}}]{verma2026quantum}%
  \BibitemOpen
  \bibfield  {author} {\bibinfo {author} {\bibfnamefont {N.}~\bibnamefont
  {Verma}}, \bibinfo {author} {\bibfnamefont {P.~J.}\ \bibnamefont {Moll}},
  \bibinfo {author} {\bibfnamefont {T.}~\bibnamefont {Holder}},\ and\ \bibinfo
  {author} {\bibfnamefont {R.}~\bibnamefont {Queiroz}},\ }\href@noop {}
  {\bibfield  {journal} {\bibinfo  {journal} {Nature Reviews Physics}\ ,\
  \bibinfo {pages} {1}} (\bibinfo {year} {2026})}\BibitemShut {NoStop}%
\bibitem [{\citenamefont {Morimoto}\ and\ \citenamefont
  {Nagaosa}(2016)}]{morimoto2016topological}%
  \BibitemOpen
  \bibfield  {author} {\bibinfo {author} {\bibfnamefont {T.}~\bibnamefont
  {Morimoto}}\ and\ \bibinfo {author} {\bibfnamefont {N.}~\bibnamefont
  {Nagaosa}},\ }\href {https://doi.org/10.1126/sciadv.1501524} {\bibfield
  {journal} {\bibinfo  {journal} {Science Advances}\ }\textbf {\bibinfo
  {volume} {2}},\ \bibinfo {pages} {e1501524} (\bibinfo {year}
  {2016})}\BibitemShut {NoStop}%
\bibitem [{\citenamefont {Peotta}\ and\ \citenamefont
  {T{\"o}rm{\"a}}(2015)}]{peotta2015superfluidity}%
  \BibitemOpen
  \bibfield  {author} {\bibinfo {author} {\bibfnamefont {S.}~\bibnamefont
  {Peotta}}\ and\ \bibinfo {author} {\bibfnamefont {P.}~\bibnamefont
  {T{\"o}rm{\"a}}},\ }\href {https://doi.org/10.1038/ncomms9944} {\bibfield
  {journal} {\bibinfo  {journal} {Nature Communications}\ }\textbf {\bibinfo
  {volume} {6}},\ \bibinfo {pages} {8944} (\bibinfo {year} {2015})}\BibitemShut
  {NoStop}%
\bibitem [{\citenamefont {Resta}\ and\ \citenamefont
  {Sorella}(1999)}]{resta1999electron}%
  \BibitemOpen
  \bibfield  {author} {\bibinfo {author} {\bibfnamefont {R.}~\bibnamefont
  {Resta}}\ and\ \bibinfo {author} {\bibfnamefont {S.}~\bibnamefont
  {Sorella}},\ }\href@noop {} {\bibfield  {journal} {\bibinfo  {journal}
  {Physical Review Letters}\ }\textbf {\bibinfo {volume} {82}},\ \bibinfo
  {pages} {370} (\bibinfo {year} {1999})}\BibitemShut {NoStop}%
\bibitem [{\citenamefont {E.K.Kudinov}(1991)}]{Kudinov91a}%
  \BibitemOpen
  \bibfield  {author} {\bibinfo {author} {\bibnamefont {E.K.Kudinov}},\
  }\href@noop {} {\bibfield  {journal} {\bibinfo  {journal} {Sov.Phys. Solid
  State}\ }\textbf {\bibinfo {volume} {33}},\ \bibinfo {pages} {1299} (\bibinfo
  {year} {1991})}\BibitemShut {NoStop}%
\bibitem [{\citenamefont {Souza}\ \emph {et~al.}(2000)\citenamefont {Souza},
  \citenamefont {Wilkens},\ and\ \citenamefont
  {Martin}}]{souza2000polarization}%
  \BibitemOpen
  \bibfield  {author} {\bibinfo {author} {\bibfnamefont {I.}~\bibnamefont
  {Souza}}, \bibinfo {author} {\bibfnamefont {T.}~\bibnamefont {Wilkens}},\
  and\ \bibinfo {author} {\bibfnamefont {R.~M.}\ \bibnamefont {Martin}},\
  }\href {https://doi.org/10.1103/PhysRevB.62.1666} {\bibfield  {journal}
  {\bibinfo  {journal} {Physical Review B}\ }\textbf {\bibinfo {volume} {62}},\
  \bibinfo {pages} {1666} (\bibinfo {year} {2000})}\BibitemShut {NoStop}%
\bibitem [{\citenamefont {Verma}\ and\ \citenamefont
  {Queiroz}(2025)}]{verma2025instantaneous}%
  \BibitemOpen
  \bibfield  {author} {\bibinfo {author} {\bibfnamefont {N.}~\bibnamefont
  {Verma}}\ and\ \bibinfo {author} {\bibfnamefont {R.}~\bibnamefont
  {Queiroz}},\ }\href@noop {} {\bibfield  {journal} {\bibinfo  {journal}
  {Proceedings of the National Academy of Sciences}\ }\textbf {\bibinfo
  {volume} {122}},\ \bibinfo {pages} {e2405837122} (\bibinfo {year}
  {2025})}\BibitemShut {NoStop}%
\bibitem [{\citenamefont {Souza}\ \emph {et~al.}(2025)\citenamefont {Souza},
  \citenamefont {Martin},\ and\ \citenamefont {Stengel}}]{souza2025optical}%
  \BibitemOpen
  \bibfield  {author} {\bibinfo {author} {\bibfnamefont {I.}~\bibnamefont
  {Souza}}, \bibinfo {author} {\bibfnamefont {R.}~\bibnamefont {Martin}},\ and\
  \bibinfo {author} {\bibfnamefont {M.}~\bibnamefont {Stengel}},\ }\href@noop
  {} {\bibfield  {journal} {\bibinfo  {journal} {SciPost Physics}\ }\textbf
  {\bibinfo {volume} {18}},\ \bibinfo {pages} {127} (\bibinfo {year}
  {2025})}\BibitemShut {NoStop}%
\bibitem [{\citenamefont {Onishi}\ and\ \citenamefont
  {Fu}(2025)}]{onishi2025quantum}%
  \BibitemOpen
  \bibfield  {author} {\bibinfo {author} {\bibfnamefont {Y.}~\bibnamefont
  {Onishi}}\ and\ \bibinfo {author} {\bibfnamefont {L.}~\bibnamefont {Fu}},\
  }\href@noop {} {\bibfield  {journal} {\bibinfo  {journal} {Physical Review
  Research}\ }\textbf {\bibinfo {volume} {7}},\ \bibinfo {pages} {023158}
  (\bibinfo {year} {2025})}\BibitemShut {NoStop}%
\bibitem [{\citenamefont {Hauke}\ \emph {et~al.}(2016)\citenamefont {Hauke},
  \citenamefont {Heyl}, \citenamefont {Tagliacozzo},\ and\ \citenamefont
  {Zoller}}]{hauke2016measuring}%
  \BibitemOpen
  \bibfield  {author} {\bibinfo {author} {\bibfnamefont {P.}~\bibnamefont
  {Hauke}}, \bibinfo {author} {\bibfnamefont {M.}~\bibnamefont {Heyl}},
  \bibinfo {author} {\bibfnamefont {L.}~\bibnamefont {Tagliacozzo}},\ and\
  \bibinfo {author} {\bibfnamefont {P.}~\bibnamefont {Zoller}},\ }\href@noop {}
  {\bibfield  {journal} {\bibinfo  {journal} {Nature Physics}\ }\textbf
  {\bibinfo {volume} {12}},\ \bibinfo {pages} {778} (\bibinfo {year}
  {2016})}\BibitemShut {NoStop}%
\bibitem [{\citenamefont {Scheie}\ \emph {et~al.}(2021)\citenamefont {Scheie},
  \citenamefont {Laurell}, \citenamefont {Samarakoon}, \citenamefont {Lake},
  \citenamefont {Nagler}, \citenamefont {Granroth}, \citenamefont {Okamoto},
  \citenamefont {Alvarez},\ and\ \citenamefont
  {Tennant}}]{scheie2021witnessing}%
  \BibitemOpen
  \bibfield  {author} {\bibinfo {author} {\bibfnamefont {A.}~\bibnamefont
  {Scheie}}, \bibinfo {author} {\bibfnamefont {P.}~\bibnamefont {Laurell}},
  \bibinfo {author} {\bibfnamefont {A.~M.}\ \bibnamefont {Samarakoon}},
  \bibinfo {author} {\bibfnamefont {B.}~\bibnamefont {Lake}}, \bibinfo {author}
  {\bibfnamefont {S.}~\bibnamefont {Nagler}}, \bibinfo {author} {\bibfnamefont
  {G.}~\bibnamefont {Granroth}}, \bibinfo {author} {\bibfnamefont
  {S.}~\bibnamefont {Okamoto}}, \bibinfo {author} {\bibfnamefont
  {G.}~\bibnamefont {Alvarez}},\ and\ \bibinfo {author} {\bibfnamefont
  {D.}~\bibnamefont {Tennant}},\ }\href@noop {} {\bibfield  {journal} {\bibinfo
   {journal} {Physical Review B}\ }\textbf {\bibinfo {volume} {103}},\ \bibinfo
  {pages} {224434} (\bibinfo {year} {2021})}\BibitemShut {NoStop}%
\bibitem [{\citenamefont {Herzfeld}(1927)}]{herzfeld1927atomic}%
  \BibitemOpen
  \bibfield  {author} {\bibinfo {author} {\bibfnamefont {K.}~\bibnamefont
  {Herzfeld}},\ }\href@noop {} {\bibfield  {journal} {\bibinfo  {journal}
  {Physical Review}\ }\textbf {\bibinfo {volume} {29}},\ \bibinfo {pages} {701}
  (\bibinfo {year} {1927})}\BibitemShut {NoStop}%
\bibitem [{\citenamefont {Brauwers}\ \emph {et~al.}(1975)\citenamefont
  {Brauwers}, \citenamefont {Evrard},\ and\ \citenamefont
  {Kartheuser}}]{brauwers1975sum}%
  \BibitemOpen
  \bibfield  {author} {\bibinfo {author} {\bibfnamefont {M.}~\bibnamefont
  {Brauwers}}, \bibinfo {author} {\bibfnamefont {R.}~\bibnamefont {Evrard}},\
  and\ \bibinfo {author} {\bibfnamefont {E.}~\bibnamefont {Kartheuser}},\
  }\href@noop {} {\bibfield  {journal} {\bibinfo  {journal} {Physical Review
  B}\ }\textbf {\bibinfo {volume} {12}},\ \bibinfo {pages} {5864} (\bibinfo
  {year} {1975})}\BibitemShut {NoStop}%
\bibitem [{\citenamefont {Kivelson}(1982)}]{kivelson1982wannier}%
  \BibitemOpen
  \bibfield  {author} {\bibinfo {author} {\bibfnamefont {S.}~\bibnamefont
  {Kivelson}},\ }\href@noop {} {\bibfield  {journal} {\bibinfo  {journal}
  {Physical Review B}\ }\textbf {\bibinfo {volume} {26}},\ \bibinfo {pages}
  {4269} (\bibinfo {year} {1982})}\BibitemShut {NoStop}%
\bibitem [{\citenamefont {Marzari}\ and\ \citenamefont
  {Vanderbilt}(1997)}]{marzari1997maximally}%
  \BibitemOpen
  \bibfield  {author} {\bibinfo {author} {\bibfnamefont {N.}~\bibnamefont
  {Marzari}}\ and\ \bibinfo {author} {\bibfnamefont {D.}~\bibnamefont
  {Vanderbilt}},\ }\href {https://doi.org/10.1103/PhysRevB.56.12847} {\bibfield
   {journal} {\bibinfo  {journal} {Physical Review B}\ }\textbf {\bibinfo
  {volume} {56}},\ \bibinfo {pages} {12847} (\bibinfo {year}
  {1997})}\BibitemShut {NoStop}%
\bibitem [{\citenamefont {Traini}(1996)}]{traini1996electric}%
  \BibitemOpen
  \bibfield  {author} {\bibinfo {author} {\bibfnamefont {M.}~\bibnamefont
  {Traini}},\ }\href@noop {} {\bibfield  {journal} {\bibinfo  {journal}
  {European Journal of Physics}\ }\textbf {\bibinfo {volume} {17}},\ \bibinfo
  {pages} {30} (\bibinfo {year} {1996})}\BibitemShut {NoStop}%
\bibitem [{\citenamefont {Mao}\ \emph {et~al.}(2025)\citenamefont {Mao},
  \citenamefont {Mendez-Valderrama},\ and\ \citenamefont
  {Chowdhury}}]{mao2025low}%
  \BibitemOpen
  \bibfield  {author} {\bibinfo {author} {\bibfnamefont {D.}~\bibnamefont
  {Mao}}, \bibinfo {author} {\bibfnamefont {J.~F.}\ \bibnamefont
  {Mendez-Valderrama}},\ and\ \bibinfo {author} {\bibfnamefont
  {D.}~\bibnamefont {Chowdhury}},\ }\href@noop {} {\bibfield  {journal}
  {\bibinfo  {journal} {Physical Review B}\ }\textbf {\bibinfo {volume}
  {112}},\ \bibinfo {pages} {075116} (\bibinfo {year} {2025})}\BibitemShut
  {NoStop}%
\bibitem [{\citenamefont {Ozawa}\ and\ \citenamefont
  {Goldman}(2019)}]{ozawa2019probing}%
  \BibitemOpen
  \bibfield  {author} {\bibinfo {author} {\bibfnamefont {T.}~\bibnamefont
  {Ozawa}}\ and\ \bibinfo {author} {\bibfnamefont {N.}~\bibnamefont
  {Goldman}},\ }\href@noop {} {\bibfield  {journal} {\bibinfo  {journal}
  {Physical Review Research}\ }\textbf {\bibinfo {volume} {1}},\ \bibinfo
  {pages} {032019} (\bibinfo {year} {2019})}\BibitemShut {NoStop}%
\bibitem [{\citenamefont {Faugno}\ and\ \citenamefont
  {Ozawa}(2026)}]{faugno2026characterizing}%
  \BibitemOpen
  \bibfield  {author} {\bibinfo {author} {\bibfnamefont {W.}~\bibnamefont
  {Faugno}}\ and\ \bibinfo {author} {\bibfnamefont {T.}~\bibnamefont {Ozawa}},\
  }\href@noop {} {\bibfield  {journal} {\bibinfo  {journal} {Physical Review
  B}\ }\textbf {\bibinfo {volume} {113}},\ \bibinfo {pages} {014306} (\bibinfo
  {year} {2026})}\BibitemShut {NoStop}%
\bibitem [{\citenamefont {Ba{\l}ut}\ \emph {et~al.}(2025)\citenamefont
  {Ba{\l}ut}, \citenamefont {Bradlyn},\ and\ \citenamefont
  {Abbamonte}}]{balut2025quantum}%
  \BibitemOpen
  \bibfield  {author} {\bibinfo {author} {\bibfnamefont {D.}~\bibnamefont
  {Ba{\l}ut}}, \bibinfo {author} {\bibfnamefont {B.}~\bibnamefont {Bradlyn}},\
  and\ \bibinfo {author} {\bibfnamefont {P.}~\bibnamefont {Abbamonte}},\
  }\href@noop {} {\bibfield  {journal} {\bibinfo  {journal} {Physical Review
  B}\ }\textbf {\bibinfo {volume} {111}},\ \bibinfo {pages} {125161} (\bibinfo
  {year} {2025})}\BibitemShut {NoStop}%
\bibitem [{\citenamefont {Ba{\l}ut}\ \emph {et~al.}(2026)\citenamefont
  {Ba{\l}ut}, \citenamefont {Bradlyn}, \citenamefont {Collins},\ and\
  \citenamefont {Abbamonte}}]{balut2026fundamental}%
  \BibitemOpen
  \bibfield  {author} {\bibinfo {author} {\bibfnamefont {D.}~\bibnamefont
  {Ba{\l}ut}}, \bibinfo {author} {\bibfnamefont {B.}~\bibnamefont {Bradlyn}},
  \bibinfo {author} {\bibfnamefont {M.~D.}\ \bibnamefont {Collins}},\ and\
  \bibinfo {author} {\bibfnamefont {P.}~\bibnamefont {Abbamonte}},\ }\href@noop
  {} {\bibfield  {journal} {\bibinfo  {journal} {arXiv preprint
  arXiv:2601.19054}\ } (\bibinfo {year} {2026})}\BibitemShut {NoStop}%
\bibitem [{\citenamefont {Kang}\ \emph {et~al.}(2025)\citenamefont {Kang},
  \citenamefont {Kim}, \citenamefont {Qian}, \citenamefont {Neves},
  \citenamefont {Ye}, \citenamefont {Jung}, \citenamefont {Puntel},
  \citenamefont {Mazzola}, \citenamefont {Fang}, \citenamefont {Jozwiak} \emph
  {et~al.}}]{kang2025measurements}%
  \BibitemOpen
  \bibfield  {author} {\bibinfo {author} {\bibfnamefont {M.}~\bibnamefont
  {Kang}}, \bibinfo {author} {\bibfnamefont {S.}~\bibnamefont {Kim}}, \bibinfo
  {author} {\bibfnamefont {Y.}~\bibnamefont {Qian}}, \bibinfo {author}
  {\bibfnamefont {P.~M.}\ \bibnamefont {Neves}}, \bibinfo {author}
  {\bibfnamefont {L.}~\bibnamefont {Ye}}, \bibinfo {author} {\bibfnamefont
  {J.}~\bibnamefont {Jung}}, \bibinfo {author} {\bibfnamefont {D.}~\bibnamefont
  {Puntel}}, \bibinfo {author} {\bibfnamefont {F.}~\bibnamefont {Mazzola}},
  \bibinfo {author} {\bibfnamefont {S.}~\bibnamefont {Fang}}, \bibinfo {author}
  {\bibfnamefont {C.}~\bibnamefont {Jozwiak}}, \emph {et~al.},\ }\href@noop {}
  {\bibfield  {journal} {\bibinfo  {journal} {Nature Physics}\ }\textbf
  {\bibinfo {volume} {21}},\ \bibinfo {pages} {110} (\bibinfo {year}
  {2025})}\BibitemShut {NoStop}%
\bibitem [{\citenamefont {Capizzi}\ \emph {et~al.}(1980)\citenamefont
  {Capizzi}, \citenamefont {Thomas}, \citenamefont {DeRosa}, \citenamefont
  {Bhatt},\ and\ \citenamefont {Rice}}]{capizzi1980observation}%
  \BibitemOpen
  \bibfield  {author} {\bibinfo {author} {\bibfnamefont {M.}~\bibnamefont
  {Capizzi}}, \bibinfo {author} {\bibfnamefont {G.}~\bibnamefont {Thomas}},
  \bibinfo {author} {\bibfnamefont {F.}~\bibnamefont {DeRosa}}, \bibinfo
  {author} {\bibfnamefont {R.}~\bibnamefont {Bhatt}},\ and\ \bibinfo {author}
  {\bibfnamefont {T.}~\bibnamefont {Rice}},\ }\href@noop {} {\bibfield
  {journal} {\bibinfo  {journal} {Physical Review Letters}\ }\textbf {\bibinfo
  {volume} {44}},\ \bibinfo {pages} {1019} (\bibinfo {year}
  {1980})}\BibitemShut {NoStop}%
\bibitem [{\citenamefont {Rosenbaum}\ \emph {et~al.}(1983)\citenamefont
  {Rosenbaum}, \citenamefont {Milligan}, \citenamefont {Paalanen},
  \citenamefont {Thomas}, \citenamefont {Bhatt},\ and\ \citenamefont
  {Lin}}]{rosenbaum1983metal}%
  \BibitemOpen
  \bibfield  {author} {\bibinfo {author} {\bibfnamefont {T.}~\bibnamefont
  {Rosenbaum}}, \bibinfo {author} {\bibfnamefont {R.}~\bibnamefont {Milligan}},
  \bibinfo {author} {\bibfnamefont {M.}~\bibnamefont {Paalanen}}, \bibinfo
  {author} {\bibfnamefont {G.}~\bibnamefont {Thomas}}, \bibinfo {author}
  {\bibfnamefont {R.~N.}\ \bibnamefont {Bhatt}},\ and\ \bibinfo {author}
  {\bibfnamefont {W.}~\bibnamefont {Lin}},\ }\href@noop {} {\bibfield
  {journal} {\bibinfo  {journal} {Phys. Rev. B}\ }\textbf {\bibinfo {volume}
  {27}},\ \bibinfo {pages} {7509} (\bibinfo {year} {1983})}\BibitemShut
  {NoStop}%
\bibitem [{\citenamefont {Helgren}\ \emph {et~al.}(2002)\citenamefont
  {Helgren}, \citenamefont {Armitage},\ and\ \citenamefont
  {Gr\"uner}}]{Helgren02a}%
  \BibitemOpen
  \bibfield  {author} {\bibinfo {author} {\bibfnamefont {E.}~\bibnamefont
  {Helgren}}, \bibinfo {author} {\bibfnamefont {N.~P.}\ \bibnamefont
  {Armitage}},\ and\ \bibinfo {author} {\bibfnamefont {G.}~\bibnamefont
  {Gr\"uner}},\ }\href {https://doi.org/10.1103/PhysRevLett.89.246601}
  {\bibfield  {journal} {\bibinfo  {journal} {Phys. Rev. Lett.}\ }\textbf
  {\bibinfo {volume} {89}},\ \bibinfo {pages} {246601} (\bibinfo {year}
  {2002})}\BibitemShut {NoStop}%
\bibitem [{\citenamefont {Helgren}\ \emph {et~al.}(2004)\citenamefont
  {Helgren}, \citenamefont {Armitage},\ and\ \citenamefont
  {Gr\"uner}}]{Helgren04a}%
  \BibitemOpen
  \bibfield  {author} {\bibinfo {author} {\bibfnamefont {E.}~\bibnamefont
  {Helgren}}, \bibinfo {author} {\bibfnamefont {N.~P.}\ \bibnamefont
  {Armitage}},\ and\ \bibinfo {author} {\bibfnamefont {G.}~\bibnamefont
  {Gr\"uner}},\ }\href {https://doi.org/10.1103/PhysRevB.69.014201} {\bibfield
  {journal} {\bibinfo  {journal} {Phys. Rev. B}\ }\textbf {\bibinfo {volume}
  {69}},\ \bibinfo {pages} {014201} (\bibinfo {year} {2004})}\BibitemShut
  {NoStop}%
\bibitem [{\citenamefont {Hering}\ \emph {et~al.}(2007)\citenamefont {Hering},
  \citenamefont {Scheffler}, \citenamefont {Dressel},\ and\ \citenamefont
  {L{\"o}hneysen}}]{hering2007signature}%
  \BibitemOpen
  \bibfield  {author} {\bibinfo {author} {\bibfnamefont {M.}~\bibnamefont
  {Hering}}, \bibinfo {author} {\bibfnamefont {M.}~\bibnamefont {Scheffler}},
  \bibinfo {author} {\bibfnamefont {M.}~\bibnamefont {Dressel}},\ and\ \bibinfo
  {author} {\bibfnamefont {H.~v.}\ \bibnamefont {L{\"o}hneysen}},\ }\href@noop
  {} {\bibfield  {journal} {\bibinfo  {journal} {Physical Review B}\ }\textbf
  {\bibinfo {volume} {75}},\ \bibinfo {pages} {205203} (\bibinfo {year}
  {2007})}\BibitemShut {NoStop}%
\bibitem [{\citenamefont {Gaymann}\ \emph {et~al.}(1993)\citenamefont
  {Gaymann}, \citenamefont {Geserich},\ and\ \citenamefont
  {L{\"o}hneysen}}]{Gayman93a}%
  \BibitemOpen
  \bibfield  {author} {\bibinfo {author} {\bibfnamefont {A.}~\bibnamefont
  {Gaymann}}, \bibinfo {author} {\bibfnamefont {H.}~\bibnamefont {Geserich}},\
  and\ \bibinfo {author} {\bibfnamefont {H.~v.}\ \bibnamefont
  {L{\"o}hneysen}},\ }\href@noop {} {\bibfield  {journal} {\bibinfo  {journal}
  {Physical review letters}\ }\textbf {\bibinfo {volume} {71}},\ \bibinfo
  {pages} {3681} (\bibinfo {year} {1993})}\BibitemShut {NoStop}%
\bibitem [{\citenamefont {Thomas}\ \emph {et~al.}(1981)\citenamefont {Thomas},
  \citenamefont {Capizzi}, \citenamefont {DeRosa}, \citenamefont {Bhatt},\ and\
  \citenamefont {Rice}}]{Thomas81a}%
  \BibitemOpen
  \bibfield  {author} {\bibinfo {author} {\bibfnamefont {G.}~\bibnamefont
  {Thomas}}, \bibinfo {author} {\bibfnamefont {M.}~\bibnamefont {Capizzi}},
  \bibinfo {author} {\bibfnamefont {F.}~\bibnamefont {DeRosa}}, \bibinfo
  {author} {\bibfnamefont {R.}~\bibnamefont {Bhatt}},\ and\ \bibinfo {author}
  {\bibfnamefont {T.}~\bibnamefont {Rice}},\ }\href@noop {} {\bibfield
  {journal} {\bibinfo  {journal} {Physical Review B}\ }\textbf {\bibinfo
  {volume} {23}},\ \bibinfo {pages} {5472} (\bibinfo {year}
  {1981})}\BibitemShut {NoStop}%
\bibitem [{\citenamefont {Smith}\ \emph {et~al.}(2017)\citenamefont {Smith},
  \citenamefont {Budi}, \citenamefont {Per}, \citenamefont {Vogt},
  \citenamefont {Drumm}, \citenamefont {Hollenberg}, \citenamefont {Cole},\
  and\ \citenamefont {Russo}}]{smith2017ab}%
  \BibitemOpen
  \bibfield  {author} {\bibinfo {author} {\bibfnamefont {J.}~\bibnamefont
  {Smith}}, \bibinfo {author} {\bibfnamefont {A.}~\bibnamefont {Budi}},
  \bibinfo {author} {\bibfnamefont {M.}~\bibnamefont {Per}}, \bibinfo {author}
  {\bibfnamefont {N.}~\bibnamefont {Vogt}}, \bibinfo {author} {\bibfnamefont
  {D.}~\bibnamefont {Drumm}}, \bibinfo {author} {\bibfnamefont
  {L.}~\bibnamefont {Hollenberg}}, \bibinfo {author} {\bibfnamefont
  {J.}~\bibnamefont {Cole}},\ and\ \bibinfo {author} {\bibfnamefont
  {S.}~\bibnamefont {Russo}},\ }\href@noop {} {\bibfield  {journal} {\bibinfo
  {journal} {Scientific reports}\ }\textbf {\bibinfo {volume} {7}},\ \bibinfo
  {pages} {6010} (\bibinfo {year} {2017})}\BibitemShut {NoStop}%
\bibitem [{\citenamefont {McMillan}(1981)}]{mcmillan1981scaling}%
  \BibitemOpen
  \bibfield  {author} {\bibinfo {author} {\bibfnamefont {W.}~\bibnamefont
  {McMillan}},\ }\href@noop {} {\bibfield  {journal} {\bibinfo  {journal}
  {Physical Review B}\ }\textbf {\bibinfo {volume} {24}},\ \bibinfo {pages}
  {2739} (\bibinfo {year} {1981})}\BibitemShut {NoStop}%
\bibitem [{\citenamefont {Kuzmenko}(2005)}]{kuzmenko2005kramers}%
  \BibitemOpen
  \bibfield  {author} {\bibinfo {author} {\bibfnamefont {A.}~\bibnamefont
  {Kuzmenko}},\ }\href@noop {} {\bibfield  {journal} {\bibinfo  {journal}
  {Review of scientific instruments}\ }\textbf {\bibinfo {volume} {76}}
  (\bibinfo {year} {2005})}\BibitemShut {NoStop}%
\bibitem [{\citenamefont {Shklovskii}\ and\ \citenamefont
  {Efros}(1981)}]{Shklovskii81a}%
  \BibitemOpen
  \bibfield  {author} {\bibinfo {author} {\bibfnamefont {B.~I.}\ \bibnamefont
  {Shklovskii}}\ and\ \bibinfo {author} {\bibfnamefont {A.~L.}\ \bibnamefont
  {Efros}},\ }\href@noop {} {\bibfield  {journal} {\bibinfo  {journal} {Sov.
  Phys. JETP}\ }\textbf {\bibinfo {volume} {54}},\ \bibinfo {pages} {218}
  (\bibinfo {year} {1981})}\BibitemShut {NoStop}%
\bibitem [{\citenamefont {Lee}\ and\ \citenamefont {Stutzmann}(2001)}]{Lee01a}%
  \BibitemOpen
  \bibfield  {author} {\bibinfo {author} {\bibfnamefont {M.}~\bibnamefont
  {Lee}}\ and\ \bibinfo {author} {\bibfnamefont {M.~L.}\ \bibnamefont
  {Stutzmann}},\ }\href {https://doi.org/10.1103/PhysRevLett.87.056402}
  {\bibfield  {journal} {\bibinfo  {journal} {Phys. Rev. Lett.}\ }\textbf
  {\bibinfo {volume} {87}},\ \bibinfo {pages} {056402} (\bibinfo {year}
  {2001})}\BibitemShut {NoStop}%
\bibitem [{\citenamefont {Tan}\ and\ \citenamefont
  {Castner}(1981)}]{tan1981piezocapacitance}%
  \BibitemOpen
  \bibfield  {author} {\bibinfo {author} {\bibfnamefont {H.}~\bibnamefont
  {Tan}}\ and\ \bibinfo {author} {\bibfnamefont {T.}~\bibnamefont {Castner}},\
  }\href@noop {} {\bibfield  {journal} {\bibinfo  {journal} {Physical Review
  B}\ }\textbf {\bibinfo {volume} {23}},\ \bibinfo {pages} {3983} (\bibinfo
  {year} {1981})}\BibitemShut {NoStop}%
\bibitem [{\citenamefont {Castner}(1980)}]{castner1980dielectric}%
  \BibitemOpen
  \bibfield  {author} {\bibinfo {author} {\bibfnamefont {T.}~\bibnamefont
  {Castner}},\ }\href@noop {} {\bibfield  {journal} {\bibinfo  {journal}
  {Philosophical Magazine B}\ }\textbf {\bibinfo {volume} {42}},\ \bibinfo
  {pages} {873} (\bibinfo {year} {1980})}\BibitemShut {NoStop}%
\bibitem [{\citenamefont {Komissarov}\ \emph {et~al.}(2024)\citenamefont
  {Komissarov}, \citenamefont {Holder},\ and\ \citenamefont
  {Queiroz}}]{komissarov2024quantum}%
  \BibitemOpen
  \bibfield  {author} {\bibinfo {author} {\bibfnamefont {I.}~\bibnamefont
  {Komissarov}}, \bibinfo {author} {\bibfnamefont {T.}~\bibnamefont {Holder}},\
  and\ \bibinfo {author} {\bibfnamefont {R.}~\bibnamefont {Queiroz}},\
  }\href@noop {} {\bibfield  {journal} {\bibinfo  {journal} {Nature
  communications}\ }\textbf {\bibinfo {volume} {15}},\ \bibinfo {pages} {4621}
  (\bibinfo {year} {2024})}\BibitemShut {NoStop}%
\bibitem [{\citenamefont {Thorsm{\o}lle}\ and\ \citenamefont
  {Armitage}(2010)}]{thorsmolle2010ultrafast}%
  \BibitemOpen
  \bibfield  {author} {\bibinfo {author} {\bibfnamefont {V.}~\bibnamefont
  {Thorsm{\o}lle}}\ and\ \bibinfo {author} {\bibfnamefont {N.}~\bibnamefont
  {Armitage}},\ }\href@noop {} {\bibfield  {journal} {\bibinfo  {journal}
  {Phys. Rev. Lett.}\ }\textbf {\bibinfo {volume} {105}},\ \bibinfo {pages}
  {086601} (\bibinfo {year} {2010})}\BibitemShut {NoStop}%
\bibitem [{\citenamefont {Mahmood}\ \emph {et~al.}(2021)\citenamefont
  {Mahmood}, \citenamefont {Chaudhuri}, \citenamefont {Gopalakrishnan},
  \citenamefont {Nandkishore},\ and\ \citenamefont
  {Armitage}}]{mahmood2021observation}%
  \BibitemOpen
  \bibfield  {author} {\bibinfo {author} {\bibfnamefont {F.}~\bibnamefont
  {Mahmood}}, \bibinfo {author} {\bibfnamefont {D.}~\bibnamefont {Chaudhuri}},
  \bibinfo {author} {\bibfnamefont {S.}~\bibnamefont {Gopalakrishnan}},
  \bibinfo {author} {\bibfnamefont {R.}~\bibnamefont {Nandkishore}},\ and\
  \bibinfo {author} {\bibfnamefont {N.}~\bibnamefont {Armitage}},\ }\href@noop
  {} {\bibfield  {journal} {\bibinfo  {journal} {Nature Physics}\ }\textbf
  {\bibinfo {volume} {17}},\ \bibinfo {pages} {627} (\bibinfo {year}
  {2021})}\BibitemShut {NoStop}%
\bibitem [{\citenamefont {Thurber}\ \emph {et~al.}(1980)\citenamefont
  {Thurber}, \citenamefont {Mattis}, \citenamefont {Liu},\ and\ \citenamefont
  {Filliben}}]{Thurber80a}%
  \BibitemOpen
  \bibfield  {author} {\bibinfo {author} {\bibfnamefont {W.~R.}\ \bibnamefont
  {Thurber}}, \bibinfo {author} {\bibfnamefont {R.~L.}\ \bibnamefont {Mattis}},
  \bibinfo {author} {\bibfnamefont {Y.~M.}\ \bibnamefont {Liu}},\ and\ \bibinfo
  {author} {\bibfnamefont {J.~J.}\ \bibnamefont {Filliben}},\ }\href
  {https://doi.org/10.1149/1.2130006} {\bibfield  {journal} {\bibinfo
  {journal} {Journal of The Electrochemical Society}\ }\textbf {\bibinfo
  {volume} {127}},\ \bibinfo {pages} {1807} (\bibinfo {year}
  {1980})}\BibitemShut {NoStop}%
\bibitem [{\citenamefont {Penn}(1962)}]{penn1962wave}%
  \BibitemOpen
  \bibfield  {author} {\bibinfo {author} {\bibfnamefont {D.~R.}\ \bibnamefont
  {Penn}},\ }\href@noop {} {\bibfield  {journal} {\bibinfo  {journal} {Physical
  review}\ }\textbf {\bibinfo {volume} {128}},\ \bibinfo {pages} {2093}
  (\bibinfo {year} {1962})}\BibitemShut {NoStop}%
\bibitem [{\citenamefont {Anderson}\ \emph {et~al.}(1958)\citenamefont
  {Anderson} \emph {et~al.}}]{anderson1958absence}%
  \BibitemOpen
  \bibfield  {author} {\bibinfo {author} {\bibfnamefont {P.~W.}\ \bibnamefont
  {Anderson}} \emph {et~al.},\ }\href@noop {} {\bibfield  {journal} {\bibinfo
  {journal} {Physical review}\ }\textbf {\bibinfo {volume} {109}},\ \bibinfo
  {pages} {1492} (\bibinfo {year} {1958})}\BibitemShut {NoStop}%
\bibitem [{\citenamefont {Shapiro}\ and\ \citenamefont
  {Abrahams}(1981)}]{shapiro1981scaling}%
  \BibitemOpen
  \bibfield  {author} {\bibinfo {author} {\bibfnamefont {B.}~\bibnamefont
  {Shapiro}}\ and\ \bibinfo {author} {\bibfnamefont {E.}~\bibnamefont
  {Abrahams}},\ }\href@noop {} {\bibfield  {journal} {\bibinfo  {journal}
  {Physical Review B}\ }\textbf {\bibinfo {volume} {24}},\ \bibinfo {pages}
  {4889} (\bibinfo {year} {1981})}\BibitemShut {NoStop}%
\bibitem [{\citenamefont {Lee}\ \emph {et~al.}(2000)\citenamefont {Lee},
  \citenamefont {Carini}, \citenamefont {Baxter}, \citenamefont {Henderson},\
  and\ \citenamefont {Gruner}}]{lee2000quantum}%
  \BibitemOpen
  \bibfield  {author} {\bibinfo {author} {\bibfnamefont {H.-L.}\ \bibnamefont
  {Lee}}, \bibinfo {author} {\bibfnamefont {J.~P.}\ \bibnamefont {Carini}},
  \bibinfo {author} {\bibfnamefont {D.~V.}\ \bibnamefont {Baxter}}, \bibinfo
  {author} {\bibfnamefont {W.}~\bibnamefont {Henderson}},\ and\ \bibinfo
  {author} {\bibfnamefont {G.}~\bibnamefont {Gruner}},\ }\href@noop {}
  {\bibfield  {journal} {\bibinfo  {journal} {Science}\ }\textbf {\bibinfo
  {volume} {287}},\ \bibinfo {pages} {633} (\bibinfo {year}
  {2000})}\BibitemShut {NoStop}%
\bibitem [{\citenamefont {Olsen}\ \emph {et~al.}(2017)\citenamefont {Olsen},
  \citenamefont {Resta},\ and\ \citenamefont {Souza}}]{olsen2017metal}%
  \BibitemOpen
  \bibfield  {author} {\bibinfo {author} {\bibfnamefont {T.}~\bibnamefont
  {Olsen}}, \bibinfo {author} {\bibfnamefont {R.}~\bibnamefont {Resta}},\ and\
  \bibinfo {author} {\bibfnamefont {I.}~\bibnamefont {Souza}},\ }\href@noop {}
  {\bibfield  {journal} {\bibinfo  {journal} {Physical Review B}\ }\textbf
  {\bibinfo {volume} {95}},\ \bibinfo {pages} {045109} (\bibinfo {year}
  {2017})}\BibitemShut {NoStop}%
\bibitem [{\citenamefont {Takeshima}(1978)}]{takeshima1978unified}%
  \BibitemOpen
  \bibfield  {author} {\bibinfo {author} {\bibfnamefont {M.}~\bibnamefont
  {Takeshima}},\ }\href@noop {} {\bibfield  {journal} {\bibinfo  {journal}
  {Physical Review B}\ }\textbf {\bibinfo {volume} {17}},\ \bibinfo {pages}
  {3996} (\bibinfo {year} {1978})}\BibitemShut {NoStop}%
\end{thebibliography}%

 \section{Acknowledgments: }
 
This work at JHU was supported by the ARO MURI ``Implementation of axion electrodynamics in topological films and device" W911NF2020166.  Instrumentation development at JHU, which made these measurements possible was supported by the Gordon and Betty Moore Foundation EPiQS Initiative Grant GBMF-9454 to NPA. We would like to thank L. Balents, D. Chowdhury, M. Heyl, P. Laurell, G. Refael, G. Reid, R Resta, R. Queiroz, A. Scheie, I. Souza, D. Vanderbilt, and N. Verma for helpful discussions.
 
 \section{Author contributions:  }
 JYY performed the experiments and the analysis with the help of XG, RB, and ND.  TFR provided the crucial samples closed to the MIQPT.  NPA directed the project.   All authors contributed to the writing and editing of the manuscript.

 \section{Additional information:} 
 
\textbf{Competing interests:} The authors declare no competing interests.

\normalsize

\setcounter{figure}{0}

\clearpage

 \textbf{SUPPLEMENTARY INFORMATION}

\bigskip

\textbf{I. Extraction of conductivity from literature data}

\bigskip

We have supplemented our data by optical spectra from Refs.~\cite{Gayman93a,Thomas81a}.  Data from Ref.~\cite{Gayman93a} was taken in a reflection geometry with a Fourier transform spectrometer Bruker IFS 113v and Kramer-Kronig transformed to ultimately generate the complex conductivity.  The real part of the conductivity for several samples spanning the 3D MIQPT at $x_c = 3.5 \times 10^{18}/\mathrm{cm}^3$ were shown in Ref.~\cite{Gayman93a}.  Note, that this data was taken at 10K which from our experiments we believe is generally not low enough temperature to reach the zero temperature limit for low frequency data near the MIQPT.  The slightly elevated values of $\ell$ as compared to the trend for the Gaymann et al. data in Fig.~\ref{lengths} in the main text may arise from this.

\begin{figure}[tbp]
\includegraphics[width=0.48\textwidth]{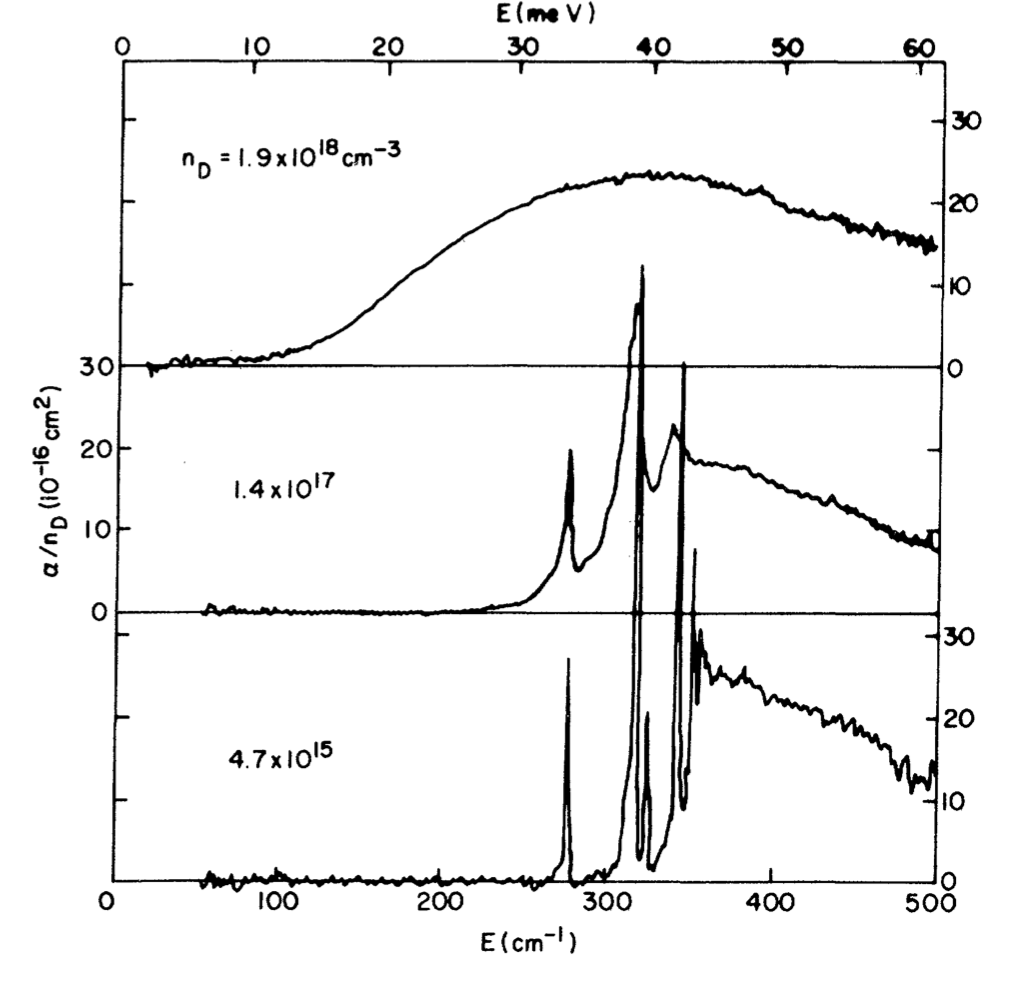}
\caption{\label{Thomas} The absorption coefficient (normalized to $x$) as a function of photon energy for three widely separated donor densities in samples of Si:P at T $<$2 K for the transmission data of Ref.~\cite{Thomas81a}.  The densities that appear in the figure are the original ones that appear in Ref.~\cite{Thomas81a}.  For use of this data in the analysis for Fig.~\ref{lengths} in the main text we have displayed the data as the same $\%$ of the critical concentrations, but recalibrated the densities by a factor of 0.95 (the approximate ratio of the critical density from the Thurber scale used in Refs.~\cite{Gayman93a,Helgren02a,Helgren04a} as compared to the critical density quoted in Ref.~\cite{Thomas81a}). }
\end{figure}

Very low density conductivity data was generated from the normalized absorption coefficients shown in Fig.~S\ref{Thomas} from Ref.~\cite{Thomas81a}.  This data was taken via transmission on samples of varying thickness. Samples were cooled to a temperature close to the surrounding liquid-He bath which is pumped to 1.1 K.  This is cold enough to be in the zero temperature limit for the measured frequency range (e.g. $\hbar \omega \gg kT$).  We calculated the real part of the conductivity from the expression $\alpha = 4\pi\sigma_1/(nc)$ using the assumption that the dielectric constant of these very low density samples is essentially the same as that of undoped silicon.  At the lowest doping ($\approx 0.1 \%$ of $x_c$) one can see in Fig.~S\ref{Thomas} numerous sharp peaks that arise largely from excitations from the $1s$ to the 2$p_\pm$ and 2$p_0$ states as well as excitons in the Mott-Hubbard gap.  As the dopant level increases one sees the peaks broaden with a notable low energy asymmetry as dopant clusters of various sizes form and tunneling is increasingly allowed between sites.

\bigskip

\textbf{II. Notes on samples}

\bigskip

Phosphorus concentrations were calibrated with the room temperature resistivity using the Thurber scale \cite{Thurber80a}.  These resistivities were remeasured from older results on this batch of samples~\cite{Helgren02a,Helgren04a} and found to be slightly systematically higher and so the dopant levels were recalibrated downward accordingly.  Also note that the Thurber scale gives a different resistivity to dopant calibration from that used in Refs~\cite{rosenbaum1983metal,Thomas81a} and the inferred critical concentrations differ accordingly.   We used the Thurber standard throughout, which gives $x_c \approx 3.51\times10^{18}$/cm$^3$, which shifts the inferred absolute doping values of the 98$\%$ and 103$\%$ samples, but not their percentage of the critical concentration.

As the interdopant interaction is exponential in their separation, the deviation from single site physics is expected when the localization length is of order the interdopant spacing.  The length scale that is relevant is not the average distance, but the most likely nearest neighbor distance.  The probability of finding the nearest neighbor dopant at a distance between $r$ and $r + dr$ follows a Poisson distribution.

\begin{align}
P_{nn}(r) dr = 4 \pi x r^2 e^{- 4 \pi x r^3/3} dr,
\end{align}
where $x$ is the dopant density.  The most probable nearest-neighbor distance from $d P_{nn} / dr = 0 $ is $\langle r_{nn} \rangle \approx 0.54 x^{-1/3}$.   This relation is plotted in Fig.~\ref{lengths}.

\bigskip
\textbf{III. Functional form for low frequency dielectric constant}
\bigskip

To analyze the correlation length on the approach to the MIQPT $\xi$ we use a form of the scaling relation for the dielectric constant motivated in part in Ref.~\cite{mcmillan1981scaling}, but here modified so that it may be valid far from the transition. We write 

\begin{align}
\epsilon = \epsilon_{si} + \epsilon_a \big(\frac{\xi}{a}\big)^{z-(d-2)}.
\end{align}
Ref.~\cite{mcmillan1981scaling} considered only the rightmost term with $d=3$, which dominates close to the transition.  We also set the microscopic prefactor as  $\epsilon_a  = 4 \pi x \alpha_D$ so that it matches the Clausius-Mossotti expression at dopant levels $x \ll x_c$ where  $\xi  \cong a$.   In the main text and below we have taken the dynamic exponent $z=3$, which is the standard $z=d$ result for a non-interacting Anderson transition.   This scaling follows from considerations of how the level spacing inside a localization volume scales e.g. $\delta E \propto \xi^{-d}$.   Previous work has found $z \approx 3$~\cite{rosenbaum1983metal}, but even aside from that we can assume the non-interacting limit in the present context because our concern is primarily in giving a {\it lower} bound on the critical length.  Considering a smaller $z$  would only serve to increase the estimate for $\xi$.

\bigskip
\textbf{IV. Bound on quantum geometric length from localization length}
\bigskip

In the spirit of Ref.~\cite{souza2025optical} we can derive a bound on the quantum geometric length ($\ell$) from the scale of the envelope of the wavefunction for an Anderson localized particle ($\xi$).  Absent any other considerations elsewhere in this manuscript these lengths cannot be the same as $\xi$ is the localization length for states at $E_F$, whereas $\ell$ is a property of the whole spectrum.  Starting from the SWM sum rule we have

\begin{align}
S_{-1} =  \frac{2}{\pi}\int_0^\infty \frac{\sigma(\omega)}{\omega}\,d\omega
&= \frac{2e^2 n \ell^2}{\hbar}.
\end{align}

We can identify a ``localization gap" $E
_L$ that is defined as the energy scale that would satisfy the SWM sum rule in a single mode approximation (e.g. if all absorptive spectral weight were concentrated at it).   $E
_L$ is then

\begin{align}
\frac{2}{\pi}\frac{1}{E_L}\int_0^\infty \sigma(\omega)\,d\omega
&= \frac{2e^2 n \ell^2}{\hbar^2}.
\label{localizationgap}
\end{align}

Next we introduce the ``Penn gap" ($E_P$)~\cite{penn1962wave}, which is the energy scale that would give the correct susceptibility in a single mode approximation.   We use the usual Kramers-Kronig relation for the low frequency susceptibility $\chi$ that is

\begin{align}
S_{-2} =  \frac{2}{\pi}\int_0^\infty \frac{\sigma(\omega)}{\omega^2}\,d\omega
&= \varepsilon_0 \chi(0) .
\end{align}
The Penn gap is defined by writing 

\begin{align}
 \frac{2}{\pi} \frac{1}{E_P^2}\int_0^\infty \sigma(\omega) d\omega
&= \frac{ \varepsilon_0   }{\hbar^2} \chi(0) . \label{PennGap} 
\end{align}

The normal f-sum rule applied to the conductivity is

\begin{align}
S_{0} =  \frac{2}{\pi} \int_0^\infty \sigma(\omega) d\omega
&= \varepsilon_0 \omega_P^2 = \frac{n e^2}{m}, 
\label{fsum}
\end{align}
where $\omega_p$ is the plasma frequency that sets the spectral weight in the low frequency conductivity, $n$ is the charge density, and $m$ is an appropriate mass.   This can be used to further evaluate the expressions above. 

The intuition behind the SWM sum rule can be found by substituting Eq.~\ref{fsum} into Eq.~\ref{localizationgap} and finding that $   \frac{1}{ 2 m } (\frac{\hbar}{\ell} )^2 = E_L$  e.g. $\frac{\hbar}{\ell}$ is an average momentum scale to make a wavepacket with energy $E_L$. $\ell$ denotes a minimum length scale that particles cannot be be localized below.  Combining Eqs.~\ref{PennGap} and~\ref{fsum} it follows that
\begin{align}
\chi = \left(\frac{\hbar\omega_p}{E_p}\right)^2 =  \left(\frac{\hbar n e^2}{m \varepsilon_0 E_p}\right)^2.
\label{susceptibility}
\end{align}

Using mathematical relations for the various moments of an integrable function, Refs.~\cite{souza2025optical,onishi2025quantum,verma2025instantaneous} showed that $E_L \geq E_P$ in general.  Using this and Eq.~\ref{localizationgap} one can infer that the quantum geometric length is bounded as

\begin{align}
\frac{2}{\pi}\frac{1}{E_p}\int_0^\infty \sigma\,d\omega
&\geq \frac{2e^2 n \ell^2}{\hbar^2} .
\end{align}

Again using the f-sum rule of Eq.~\ref{fsum} we have

\begin{align}
\frac{ \varepsilon_0 \omega_p^2}{ E_p} 
= \frac{n e^2}{m} \frac{1 }{  E_P} &\geq 2 \frac{n  e^2  \ell^2}{\hbar^2}. 
\end{align}

Substituting into Eq.~\ref{susceptibility}, we can write

\begin{align}
\frac{ \hbar }{2 e }  \sqrt{  \frac{\varepsilon_0 \chi }{m n } } \geq \ell^2  .
\end{align}
This is the strong upper bound of Ref.~\cite{souza2025optical} that for hydrogenic orbitals is almost an equality.  Now using the scaling relation introduced in the main text and above for the susceptibility as a function of the correlation length\footnote{This is equivalent to taking the non-interacting theory of $z=3$.} we get is

\begin{align}
\chi = 4 \pi x \alpha_D  \left( \frac{\xi}{a} \right )^2,
\end{align}
where $a = \frac{4 \pi \epsilon_{si} \varepsilon_{0} \hbar^2}{e^2 m} $ the Bohr radius of a P dopant electron in Si, $\alpha_D = \frac{9}{2} a^3 \epsilon_{si}$ the polarizability volume, we can give a strong bound on the quantum geometric length from the localization length.   It is

\begin{align}
\frac{3}{2 \sqrt{2}}  \frac{\xi}{a} \geq \left (\frac{\ell}{a} \right )^2 .
\end{align}

This relation is the main result of this section.

\bigskip

\textbf{V. The quantum geometric length at the metal-insulator transition }
\bigskip

It is interesting to ask what length scale plays the role of the critical length for a many-body disordered system as it approaches its MIQPT.   The usual single particle localization length $\xi$, which plays this role for the Anderson MIQPT of non-interacting fermions~\cite{anderson1958absence} and corresponds to the scale of the exponential tail of the single-particle wavefunction at $E_F$, has no obvious relevance to the many-body case.  A candidate is this quantum geometric localization length~\cite{resta1999electron} that is defined for a many-body ground state $|\Psi_0\rangle$ as
\begin{equation}
\ell^2 = -\frac{L^2}{2\pi^2} \ln \left|\left\langle \Psi_0 \left| e^{i\frac{2\pi}{L}\hat{X}} \right| \Psi_0 \right\rangle\right|^2,
\end{equation}
where $\hat{X} = \sum_i x_i$ is the total position operator. It is finite in the insulating phase and infinite in the metallic phase.   If it can play the role of a critical length in the many-body case, it should be parametrically related to $\xi$ in the non-interacting Anderson insulator limit.   Here we show  it is unrelated to $\xi$ and shows no critical behavior on the approach to the 3D MIQPT from the insulating side.

Via the Souza-Wilkens-Martin (SWM) sum rule~\cite{souza2000polarization}, $\ell$ is  related to the optical conductivity as
\begin{equation}
\ell^2 = \frac{\hbar}{\pi e^2 n_e} \int_0^\infty \frac{\mathrm{Re}\,\sigma(\omega)}{\omega}\,d\omega.
\label{SWMAppendix}
\end{equation}
This is distinct from the $1/\omega^2$-weighted integral that gives the static dielectric constant via Kramers-Kronig that is
\begin{equation}
\varepsilon(0) - 1 \propto \int_0^\infty \frac{\mathrm{Re}\,\sigma(\omega)}{\omega^2}\,d\omega.
\end{equation}

\begin{figure}[ht]
\centering
\includegraphics[width=1\columnwidth]{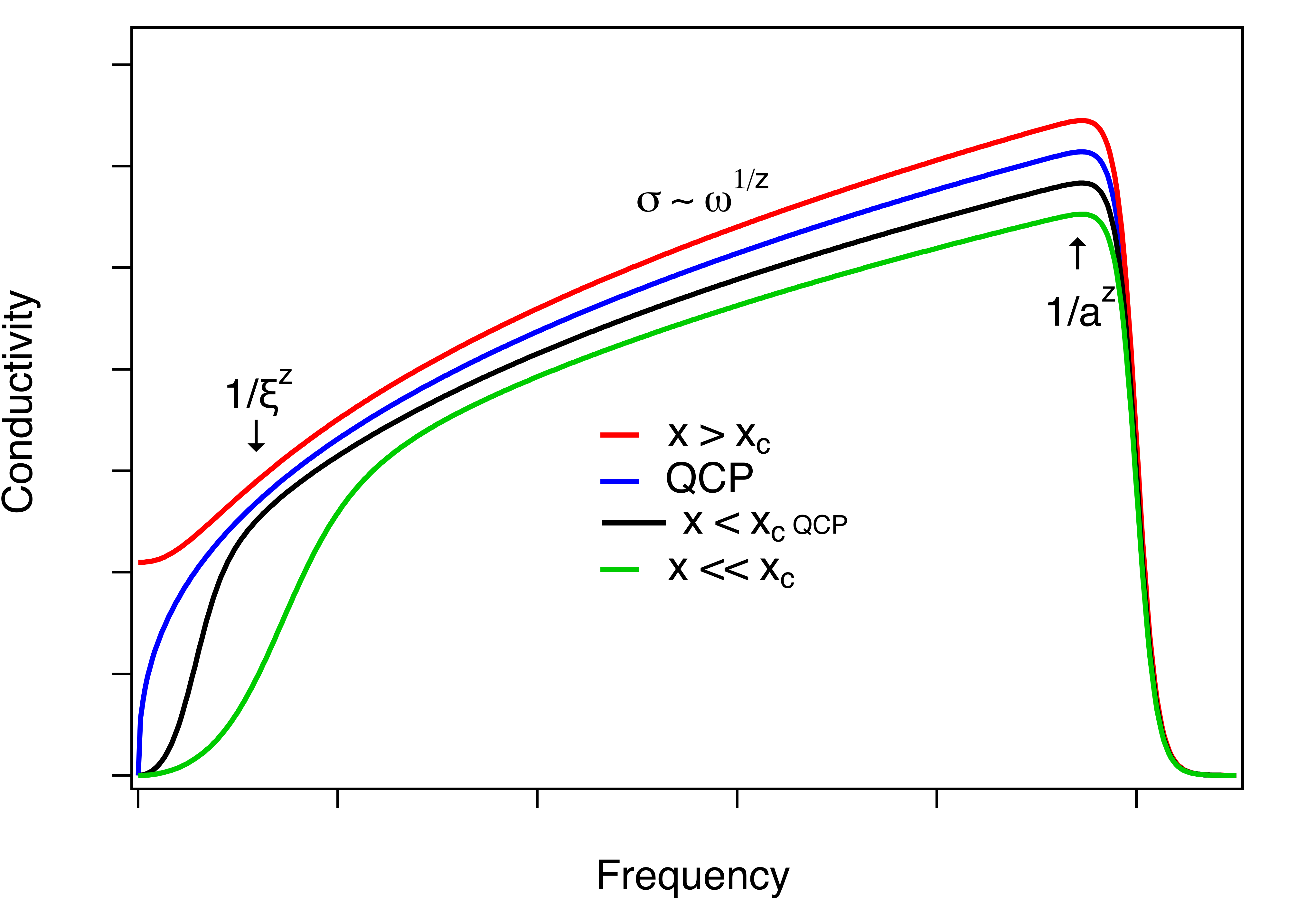}
\caption{Schematic of the optical conductivity for critical concentrations up to and through the MIQPT. } 
\label{QCPconductivity}
\end{figure}

Near a MIQPT critical point~\cite{shapiro1981scaling,lee2000quantum}, the optical conductivity obeys the scaling form
\begin{equation}
\mathrm{Re}\,\sigma(\omega) \sim \xi^{-(d-2)} F(\omega \xi^z),
\label{scalingform}
\end{equation}
where $\xi \sim |x - x_c|^{-\nu}$ is the correlation length, $z$ is the dynamical exponent, and $F(u)$ is a universal scaling function. Near the critical point and at high frequencies the conductivity will be insensitive to the proximity to the transition so the functional dependence of $\sigma$ on $\xi$ will drop out for large $\omega$.  This fixes the large-$u$ behavior of $F(u) \sim u^{(d-2)/z}$ giving $\sigma \sim \omega^{\frac{d-2}{z}}$ at large $\omega$ near the MIQPT.  One expects this dependence up to a UV cutoff that we take to be given by a weakly doping dependent short distance scale.   Its corresponding frequency is $1/a^z$.  Note that the conductivity must decrease at least as fast as $1/\omega^2$ above some frequency as required by the usual $f$-sum rule.  Hence we expect that even in the absence of the clean cutoff as shown shown in Fig.~S\ref{QCPconductivity} and with a more complicated lineshape than shown here, there are only small non-universal contributions to Eq.~\ref{SWMAppendix} from frequencies at and above the scale that we parametrize as $1/a^z$.

Substituting the scaling form of Eq.~\ref{scalingform} into the SWM integral with $u = \omega\xi^z$ we get
\begin{equation}
\ell^2 \sim \xi^{-(d-2)} \int_{u_\text{min}}^{u_\text{max}} \frac{F(u)}{u}\,du.
\end{equation}
The scaled UV cutoff is $u_\text{max} \sim (\xi/a)^z$. With $F(u) \sim u^{\frac{d-2}{z}}$ at large $u$ and $d>2$ the integral is dominated by its UV limit and is

\begin{equation}
\int^{(\xi/a)^z} u^{\frac{d-2}{z} - 1}\,du \sim \left(\xi/a\right)^{d-2}.
\end{equation}
Therefore
\begin{equation}
\ell^2 \sim \xi^{-(d-2)} \cdot \xi^{d-2} = \mathrm{const}.
\end{equation}
 
Thus we can see that the $1/\omega$ weighting in the SWM rule is insufficient to make it sensitive to the low-frequency spectral changes that define the MIQPT as shown schematically in Fig.~S\ref{QCPconductivity}.  In contrast, the high-frequency conductivity that controls the integral evolves smoothly through the transition. Consequently, $\ell$ does not diverge continuously as the 3D MIQPT is approached from the insulating side. Instead it remains finite and jumps discontinuously to infinity at the transition, where the metallic DC contribution appears and makes the IR contribution diverge.  This behavior is consistent with the numerical study of Ref.~\cite{olsen2017metal} that implied that $\ell$ and $\xi$ are unrelated in the vicinity of the MIQPT.  Notably, the behavior of $\ell$ contrasts with $\varepsilon(0)$, whose $1/\omega^2$-weighted integral is IR dominated and therefore sensitive to low frequency shifts of the spectral weight on the approach to the MIQPT, leading to a continuous divergence.

Note that the behavior of $\ell$ at the MIQPT -- whether it diverges continuously or jumps -- is controlled by whether the SWM integral is IR or UV dominated. This is in turn determined by the sign of $(d-2)/z$, which fixes the large-$u$ behavior of the scaling function $F(u)$ through the requirement that $\mathrm{Re}\,\sigma(\omega) \sim \omega^{(d-2)/z}$ at the critical point.  This scaling is then consistent with the result of Resta and Sorella \cite{resta1999electron}, who showed that $\ell$ \emph{diverges} at the 1D Mott transition.   For $d < 2$, $F(u) \sim u^{(d-2)/z}$ is a \emph{decreasing} function at large $u$ and the SWM integral is IR dominated, which means that $\ell$ diverges continuously at the MIQPT.  Of course for an insulator such a frequency dependence of the scaling function only makes sense if there is also a critical lower gap scale $\Delta$, which goes to zero at low frequency, which provides the lower cutoff of the integral.

\bigskip
\textbf{VI. Cumulative First Negative Sum Rule}
\bigskip

\begin{figure}[ht]
\centering
\includegraphics[width=1\columnwidth]{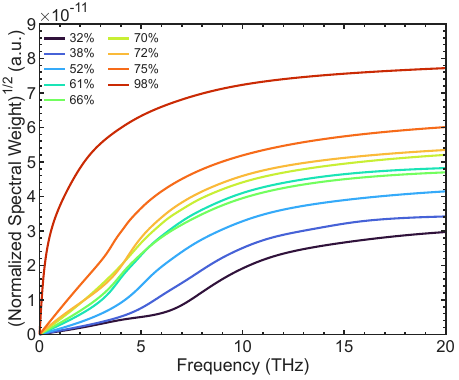}
\caption{Normalized cumulative sum rule showing the convergence of SWM integral. Note that as the doping increases, the rate of convergence becomes larger. Regardless, all of the integrals have converged to their high frequency asymptotic value well below the detection range of 200 THz.} 
\label{CumulativeSum}
\end{figure}

As shown in the previous section, the SWM integral near the 3D MIQPT is expected to be UV dominated. Here, we examine whether this behavior persists in the real MIQPT system Si:P by evaluating the SWM sum rule with a running upper cutoff.   We define
\begin{equation}
\mathrm{Normalized\ Spectral\ Weight} \propto \frac{1}{n_e} \int_0^\omega \frac{\mathrm{Re}\,\sigma(\omega)}{\omega}\,d\omega.
\end{equation}
We plot in Fig.~S\ref{CumulativeSum} the square root of this quantity as we are primarily interested in the quantum geometric length $\ell$ which is proportional to the square root of the normalized spectral weight.

As $(\mathrm{ spectral\ weight})^{1/2}$ is directly proportional to $\ell$, the curves illustrate how increasing frequency progressively contributes to the measure of the quantum geometric length. Although the $(1/\omega)$ factor does relatively enhance the low-frequency contribution, 
$\ell$ is primarily dominated by the finite-frequency peak in $\sigma_1(\omega)$. As the doping increases, the enhanced low-frequency response leads to a saturation of the weight at lower and lower frequencies.  This is a consequence of critical slowing down near the MIQPT.  For all dopings the spectral weight saturates below 20 THz, well below the experimental detection limit of 200 THz. For the 98$\%$ sample, where the low-frequency contribution is expected to be strongest, saturation occurs as low as 12 THz. 

\begin{figure}[ht]
\centering
\includegraphics[width=1\columnwidth]{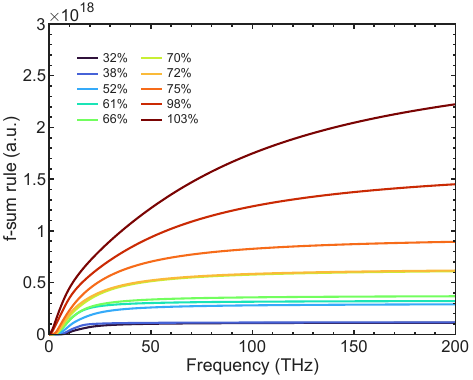}
\caption{Value of cumulative f-sum rule from integral of the optical conductivity.} 
\label{CumulativeFSum}
\end{figure}

In Fig.~S\ref{CumulativeFSum} we plot the cumulative $f$-sum integral of the conductivity itself e.g the left hand side of

\begin{equation}
 \int_0^\omega \mathrm{Re}\,\sigma(\omega) \;d\omega  =  \frac{\pi}{2} \frac{n(\omega)e^2}{m}.
 \label{cumulative}
\end{equation}

\begin{figure}[ht]
\centering
\includegraphics[width=1\columnwidth]{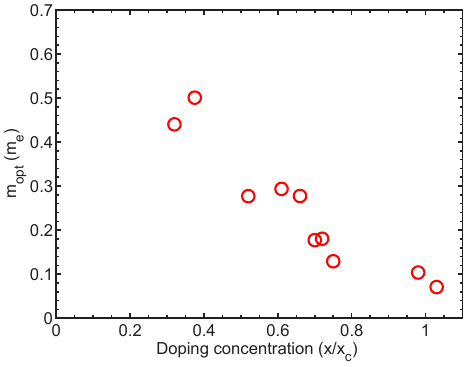}
\caption{Optical mass calculated from the optical conductivity $f$-sum rule.  } 
\label{opticalmass}
\end{figure}

In Fig.~S\ref{opticalmass} we plot the mass extracted from the high frequency limit of the right hand side of Eq.~\ref{cumulative} using $x=n$.   The optical mass decreases as the doping concentration increases from 0.55 $m_e$ to approximately 0.1 $m_e$ at $x_c$.  A similar albeit weaker dependence was reported in earlier room temperature optical studies from Ref.~\cite{Gayman93a}.

\setcounter{figure}{0}
\bigskip
\textbf{VII. Quantum geometric contribution to the Clausius-Mossotti radius}
\bigskip

As discussed in the main text we believe that we can understand the ``puffing" of the local orbitals through a quantum geometry contribution to the polarizability volume.  There are at least two microscopically distinct interactions that couple nearby donors.  These two channels contribute differently to intersite hybridization and drive the different dependencies of $\xi$ and $\ell$.

A first channel is single-particle hopping, which is a one-body term that -- considering two hydrogenic orbitals $A$ and $B$ --  goes as $t\,|1s_A\rangle\langle 1s_B|$ and hybridizes the $1s$ orbitals on adjacent sites into bonding and antibonding combinations e.g. $|\pm\rangle = \frac{|1s_A \rangle \pm  | 1s_B\rangle}{\sqrt{2}}$. The resulting dipole-active transition carries a large intersite transition dipole $\langle +|\hat{x}|-\rangle \simeq R/2$.  As this channel also transports charge in the multisite case it is ultimately responsible for building the extended envelope $\xi$, and drives the MIQPT.  This channel may lead to no polarizability enhancement as Ref.~\cite{takeshima1978unified} showed that such a ``molecule" of two hydrogenic orbitals has a smaller susceptibility than 2 such orbitals separately.

A second channel is a Coulomb dipole-dipole (van der Waals) interaction that does lead to polarizability enhancement.  This is a two-body term $V_{dd}\sim e^2 x_A x_B/R^3$ that admixes doubly excited configurations e.g. $|2p_A\,2p_B\rangle$ and by parity, only similar doubly excited states, into the ground state, since $\langle 1s|x|1s\rangle = 0$ forbids exciting one site while leaving the other in its ground state. As this channel is Coulomb only, it transports no charge, and its transition dipole is set by the atomic dipole ($\sim a$) rather than by $R$.

These two channels enter the quantum geometry differently, and this may underlie the contrasting behavior of $\ell$ and the correlation length $\xi$. It is also where the role of entanglement may appear explicitly.  Consider the case of two isolated sites.  With the total position operator $\hat{X} = x_A + x_B$, the ground-state quantum metric
$\ell^2 \propto \langle \hat{X}^2\rangle$
decomposes into single-site (diagonal) and intersite (off-diagonal) parts e.g. $ \ell^2 \propto \langle x_A^2\rangle + \langle x_B^2\rangle + 2\langle x_A x_B\rangle$.  A product (separable) ground state retains only the diagonal terms; the cross term will be nonzero only for an entangled state. The dipole-dipole channel is precisely such a off-diagonal, non-separable piece and it gives the part of the quantum Fisher information that exceeds the ``separable bound", i.e. the intersite entanglement contribution to $\ell$. Because it is pure correlation with no charge transport, it manifests in the many-body metric $\ell$ and we expect it to be largely invisible in the one-body correlation length $\xi$: the statement that the dipole-dipole chanell  feeds $\ell$ but not $\xi$ is equivalent to the statement that intersite entanglement appears in this quantum metric but not in a single-particle localization length.

A minimal model that describes this behavior is two two-level systems $A$ and $B$ with levels $|g\rangle$ and $|e\rangle$.  With gap $\Delta$ and transition dipole $d = \langle g|\hat{x}|e\rangle$ 
the two-level static polarizability is $\alpha = 2e^2 d^2/\Delta$. For a separation
$R\,\hat{z}$ the longitudinal dipole-dipole coupling is
\begin{equation}
  V_{dd} = -\frac{2e^2}{R^3}\, z_A z_B .
\end{equation}
This connects $|gg\rangle$ {\it only} to $|ee\rangle$, with
$\langle ee|V_{dd}|gg\rangle = -2e^2 d^2/R^3$, giving to leading order the entangled ground state
\begin{equation}
  |\Psi_0\rangle \simeq |gg\rangle + \lambda\,|ee\rangle ,
  \qquad \lambda = \frac{e^2 d^2}{ R^3 \Delta} .
\end{equation}
The single-site metric is unchanged at leading order,
$\langle z_A^2\rangle = \langle z_B^2\rangle = d^2$, while the off-diagonal
metric is generated entirely by the admixture.   It is
\begin{equation}
  \langle z_A z_B\rangle = 2\lambda d^2 = \frac{\alpha\, d^2}{R^3} .
\end{equation}
The longitudinal quantum metric is therefore enhanced as
\begin{equation}
  g_{zz} = \big\langle (z_A+z_B)^2 \big\rangle
         = 2d^2\!\left(1 + \frac{\alpha}{R^3}\right),
\end{equation}
so the fractional puffing of $\ell^2$ is $\Delta\ell^2/\ell^2 = \alpha/R^3 \sim 4 \pi \alpha x /3 $.   

Note that this is only the longitudinal contribution to the metric.  For isotropic dipoles the full interaction tensor is traceless as there is also a transverse contribution that will cancel the longitudinal part in an isotropic average.  Therefore entanglement driven by the simplest dipolar mixing cannot be the whole story.  How it breaks down, manifests at higher order, or is augmented via hopping are interesting topics of future study.  It would also be interesting to understand if ionized donors that should be prevalent near the MIQPT, which break the inversion of neighboring sites and allow mixing in of $|eg\rangle $-type configurations are an appreciable contribution to changes in the metric near the transition.

\end{document}